\documentclass[aps,prb,numerical,superscriptaddress,showpacs,showkeys,reprint,byrevtex]{revtex4-1} 

\usepackage{bm}
\usepackage{amsmath}
\usepackage{amsfonts}
\usepackage{amssymb}
\usepackage{graphicx}
\usepackage{setspace}

\begin{document}

\title{Stacking-dependent energetics and electronic structure of ultrathin polymorphic V$_2$VI$_3$ topological insulator nanofilms}

\author{Can Li}
 \thanks{These authors contributed equally to this work.}
 \affiliation{Global E3 Institute, Department of Materials Science and Engineering, Yonsei University, Seoul, Korea}
 \affiliation{Department of Materials Science and Engineering, China Jiliang University, Zhejiang, China}
\author{Torben Winzer}
 \thanks{These authors contributed equally to this work.}
 \affiliation{Global E3 Institute, Department of Materials Science and Engineering, Yonsei University, Seoul, Korea}
\author{Aron Walsh}
 \affiliation{Centre for Sustainable Chemical Technologies, Department of Chemistry, University of Bath, Bath, UK}
  \affiliation{Global E3 Institute, Department of Materials Science and Engineering, Yonsei University, Seoul, Korea}
\author{Binghai Yan}
 \affiliation{Max Planck Institute for Chemical Physics of Solids, N\"othnitzerStr. 40, 01187 Dresden, Germany}
  \affiliation{Max Planck Institute for Physics of Complex Systems, N\"othnitzerStr. 38, 01187 Dresden, Germany}
\author{Catherine Stampfl}
 \affiliation{School of Physics, University of Sydney, NSW 2006, Australia}
 \affiliation{Global E3 Institute, Department of Materials Science and Engineering, Yonsei University, Seoul, Korea}
\author{Aloysius Soon}
\email[Corresponding author; e-mail: ]{aloysius.soon@yonsei.ac.kr}
 \affiliation{Global E3 Institute, Department of Materials Science and Engineering, Yonsei University, Seoul, Korea}
 
\date{\today}

\begin{abstract}
Topological insulators represent a paradigm shift in surface physics. The most extensively studied Bi$_2$Se$_3$-type topological insulators exhibit layered structures, wherein neighboring layers are weakly bonded by van der Waals interactions. Using first principles density-functional theory calculations, we investigate the impact of the stacking sequence on the energetics and band structure properties of three polymorphs of Bi$_2$Se$_3$, Bi$_2$Te$_3$, and Sb$_2$Te$_3$. Considering their ultrathin films up to 6 nm as a function of its layer thickness, the overall dispersion of the band structure is found to be insensitive to the stacking sequence, while the band gap is highly sensitive, which may also affect the critical thickness for the onset of the topologically nontrivial phase. Our calculations are consistent with both experimental and theoretical results, where available. We further investigate tribological layer slippage, where we find a relatively low energy barrier between two of the considered structures. Both the stacking-dependent band gap and low slippage energy barriers, suggest that polymorphic stacking modification may offer an alternative route for controlling the properties of this new state of matter.
\end{abstract}


\maketitle

Topological insulators (TIs) have remarkable electronic properties since the role of relativistic interactions \{e.g. spin-orbit coupling (SOC)\} is fundamentally different from conventional insulators and semiconductors.\cite{Bernevig2006,Hasan2010,Moore2010,Qi2011} TIs combine an insulating band gap in the bulk of the material with conductive surface states that are protected by time-reversal symmetry.\cite{Fu2007,Hsieh2008} The topological behavior has been theoretically predicted and experimentally observed in a variety of systems,\cite{Yan2012} such as HgTe quantum wells,\cite{Bernevig2006} the Bi$_2$Se$_3$ family of compounds,\cite{Zhang2009,Wang2011,Chen2011,Scanlon2012} Heusler compounds,\cite{Chadov2010,Lin2010,Li2011} pyrochlores,\cite{Guo2009} Kondo insulators,\cite{Dzero2010} and thallium-based ternary chalcogenides.\cite{Eremeev2011} Besides the fundamental research in condensed matter physics, TIs have great potential to impact multiple areas of application (e.g. electronic, optoelectronic, and spintronic materials, thermoelectric materials, phase-change-memory and catalytic chemistry).\cite{Chen2011a,Kong2011}

To effectively explore the surface conductivity of TIs, ultrathin films with a large surface-to-volume ratio provide attractive systems for transport studies, which are highly relevant for electronic device applications.\cite{Chen2011a} For this purpose, the V$_2$VI$_3$ compounds Bi$_2$Se$_3$, Bi$_2$Te$_3$, and Sb$_2$Te$_3$ are a good choice owing to their layered rhombohedral crystal structure with space group {\it R-3m}. The structure contains five atomic layers as a basic unit, denoted as a quintuple layer (QL). There is strong chemical bonding within a QL, with weak van der Waals (vdW) interactions between different QLs. V$_2$VI$_3$ compounds can be easily grown as two-dimensional thin films by molecular beam epitaxy.\cite{Chen2011,Li2010} The two-dimensional nanostructures of Bi$_2$Se$_3$, Bi$_2$Te$_3$ and Sb$_2$Te$_3$ have been researched both theoretically and experimentally.\cite{Wang2011,Zhang2010,Liu2010,Jiang2012} Their electronic structure depends on the thickness of the film: Above a critical thickness ({\it D}), films will transform from a normal insulator (NI) to a TI.\cite{Liu2010} For the Bi$_2$Se$_3$ system, {\it D} = 6\,QL (about 6\,nm),\cite{Zhang2010,Yazyev2010} while {\it D} = 2 or 3\,QL (about 2 or 3\,nm) for Bi$_2$Te$_3$\cite{Wang2011,Yazyev2010,Park2010} and 4\,QL (about\,4 nm) for Sb$_2$Te$_3$.\cite{Jiang2012} Below the critical thickness, a surface band gap opens, due to hybridization of overlapping surface state wavefunctions.\cite{Zhang2010} Thus, a minimum film thickness is suggested for topological electronic device applications.

Although the critical thicknesses of Bi$_2$Se$_3$, Bi$_2$Te$_3$ and Sb$_2$Te$_3$ have been studied in theory and experiment, the effect of pressure and stress, which are critical factors in real environments, have not been clearly explored.\cite{Liu2011,Young2011,Zhang2011}. Liu {\it et al.} have indicated that uniaxial strain in the $\langle$111$\rangle$ direction is an important parameter for influencing the topological insulating phase and the direct band gap of Bi$_2$Se$_3$ at the $\Gamma$ point.\cite{Liu2010,Liu2011} However, besides uniaxial strain, shear strain could also be important in applications for ultrathin layered material: such as graphite,\cite{Lebedeva2011} boron nitride,\cite{Liu2003} and V$_2$VI$_3$ compounds. The weak van der Waals force between these layers leads to the possibility of layer slippage and polytypism. When layered materials are used as mechanical components of nano-devices, the properties of friction are extremely important. The nanotribologies of bilayer graphene and boron nitride layers have been intensively researched; the results show that the associated energy barriers are extremely low.\cite{Lebedeva2011,Liu2003,Popov2011} The influence of layer sequencing and slippage in V$_2$VI$_3$ TI compounds, which directly relates to their usage in electronic device applications, is the subject of this study.

\begin{figure}[tb!]
\center
\includegraphics[width=0.50\textwidth]{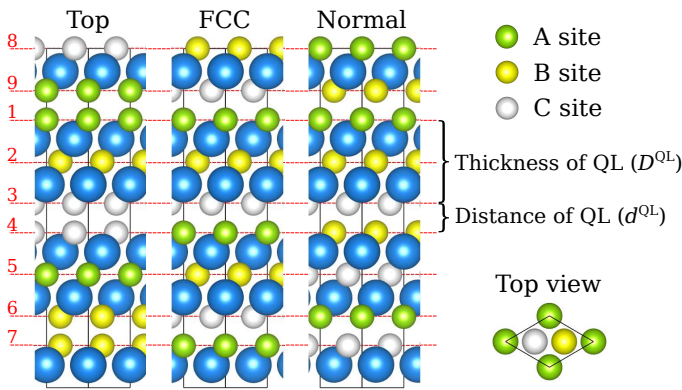}
\caption{(Color online) The atomic structures of V$_2$VI$_3$ polymorphs -- the {\it top} {(ABC-CAB-BCA)}, {\it fcc} {(ABC-ABC-ABC)} and {\it normal} {(ABC-BCA-CAB)} stacking sequence. Large spheres (blue) are used to represent the group V atoms, while the small ones (green, yellow, and white) for group VI atoms. The stacking sequence is labeled by the group VI atomic layer according to the numbers (given in red).}
\label{fig1}
\end{figure}

In these V$_2$VI$_3$ compounds, there are three QLs in each bulk conventional hexagonal unit cell, as shown in Fig.\,\ref{fig1}. There are also three unique lattice sites (A, B and C). Different polymorphic stacking types are labeled by the group VI atoms in each QL. Thus, the {\it normal} structure has a {ABC-BCA-CAB} stacking sequence, whereas the {ABC-CAB-BCA} {\it top}-like stacking (e.g. A-on-A, C-on-C etc.) is denoted as the {\it top} structure and a {ABC-ABC-ABC} {\it fcc}-like stacking will then correspond to the {\it fcc} structure. Due to the weak interaction between these QLs and the strong covalent bonding within each QL, when a shear stress is applied perpendicular to the {\it z} direction, inter-QLs could easily move or experience a mechanical slip. X-ray diffraction experiments for V$_2$VI$_3$ compounds confirm the existence of the {\it normal} structure, while the other polymorphic {\it fcc} and {\it top} structures have not yet been experimentally determined. However, if formed predominately near/at the surface, alternative polymorphic stacking sequences may not give rise to an appreciable difference in their diffraction patterns.

In this paper, we employ density-functional theory (DFT) calculations (with SOC treated explicitly) to study the effects of polymorphic stacking and planar slippage on the electronic structures and the critical thickness of the polymorphs of V$_2$VI$_3$ (namely, Bi$_2$Se$_3$, Bi$_2$Te$_3$ and Sb$_2$Te$_3$) ultrathin nanofilms as a function of film thickness.

\section{Methodology and computational approach}

All DFT calculations, including geometry relaxation and electronic structure, are performed on the basis of the projector augmented wave method\cite{Kresse1999} implemented in the Vienna {\it Ab Initio} Simulation Package (VASP) code.\cite{Kresse1996} The exchange-correlation functional used is the generalized gradient approximation (GGA) due to Perdew, Burke, and Ernzerhof (PBE),\cite{Perdew1996,Scheidemantel2003} including scalar-relativistic effects in addition to SOC. The latter is known to be of great importance in accounting for the topologically protected surface states in TIs. The kinetic energy cutoff of electron wavefunctions is set to 500\,eV and a {\bf k}-point sampling of 12$\times$12$\times$1 for all films was found to be converged. A vacuum region of 20\,{\AA} is used to avoid spurious interactions between repeating slabs. Both the shape and size of the unit cell and the relative atomic positions are relaxed with a force tolerance of 0.01\,eV/{\AA}. The dispersion-corrected DFT approach due to Grimme \textit{et al.} (DFT$+$D2) has been used in which long-range dispersion interactions are empirically described by a pair-potential of the $C_{6}/R_{0}$ form.\cite{Grimme2010} The Grimme-D2 coefficients are obtained from values tabulated in terms of the chemical identity of the atoms: $C_{6}$ = 63.565, 38.459, 12.643 and 31.750; $R_{0}$ = 1.725, 1.710, 1.610 and 1.720 for Bi, Sb, Se and Te atoms, respectively.\cite{DFTD3}

In addition, hybrid DFT calculations at the level of HSE06$+$SOC for Bi$_2$Se$_3$ nanofilms have been used to study the electronic band structure. This is because the HSE06 hybrid functional typically presents a considerable improvement over semi-local density-functionals for the description of the band gaps of solid-state systems.\cite{Franchini2005,Krukau2006} This, in turn, is compared to available results using the van der Waals density functional with Cooper's exchange (vdW-DF$^{\rm C09}_{x}$)\cite{Luo2012} and $GW$\cite{Yazyev2012} calculations in the literature.

The QL-QL interaction energies of ultrathin nanofilms, $E^{\rm QL}$ from 1\,QL to 6\,QLs have been calculated using:
\begin{equation}
E^{\rm QL}=\frac{E^{\rm sys}-N^{\rm QL}E^{\rm 1QL}}{N^{\rm QL}} \quad,
\label{eq1}
\end{equation}                                                     
where $E^{\rm sys}$ and $E^{\rm 1QL}$ are the total energy of the system in question and that of 1\,QL, respectively,  and $N^{\rm{QL}}$ is the number of QLs in that system. Since the base unit is 1\,QL, $E^{\rm{QL}}$ denotes the relative thermodynamic stability of the ultrathin film stacked per QL. In addition, to investigate the planar slippage energy barriers for these V$_2$VI$_3$ polymorphs, we employ the climbing-image nudged elastic band (CI-NEB) method at the level of PBE$+$D2.\cite{Henkelman2000}

\section{Results and discussion}
\subsection{Bulk structures of V$_2$VI$_3$ compounds}
The crystal structures for three different polymorphic stacking sequences: {\it normal}, {\it fcc} and {\it top} are shown in Fig.\,\ref{fig1}, with the optimized lattice constants listed in Tab.\,\ref{tab1}. The DFT+D2 optimized bulk structures are found to be in good agreement with the available experimental results\cite{Wyckoff1964} and the vdW-DF$^{\rm C09}_{x}$ values,\cite{Luo2012} and are much closer to these results than previously calculated theoretical results (which do not include vdW corrections).\cite{Wang2007} This signals that the DFT$+$D2 approach is adequate for describing these weakly-bonded layered systems. Cohesive energies of the bulk V$_2$VI$_3$ polymorphs are calculated and we find that all bulk structures yield a negative value, i.e. stable with respect to their corresponding atomic energies (See Tab.\,\ref{tab1}). The {\it normal} structure is the most favored bulk polymorph, while the {\it top} structure is least stable. However, for Bi$_2$Te$_3$ and Sb$_2$Te$_3$ the differences between {\it normal}, {\it fcc}, and {\it top} structures are marginal. Only the {\it normal} stacking of Bi$_2$Se$_3$ is notably more stable compared to the other polymorphs. We have calculated the electronic band structures of bulk {\it normal} Bi$_2$Se$_3$ with PBE$+$SOC and HSE$+$SOC. The band gap energy, $E_{\rm g}$ of bulk Bi$_2$Se$_3$ is calculated to be 0.36\,eV for PBE$+$SOC and 0.28\,eV for HSE06$+$SOC, respectively. Surprisingly, the PBE$+$SOC $E_{\rm g}$ seems to be in a better agreement with other theoretical reports (0.30\,eV) and experimental data (0.35\,eV) than the slightly underestimated HSE06$+$SOC value.\cite{Zhang2009,Black1957,Vidal2011} We note that for Bi$_2$Se$_3$ $GW$ corrections are known to change the character of the band gap from an indirect to a direct one.\cite{Yazyev2012,Nechaev2013}

\begin{table*}[htb!]
\centering
\caption{Lattice constants, electronic band gap, and cohesive energy of bulk V$_2$VI$_3$ compounds. The lattice constants, $a$ and $c$, and cohesive energy, $E_{\rm coh}$ are calculated using PBE$+$D2$+$SOC and are reported in {\AA} and eV, respectively. The electronic band gap, $E_{\rm g}$ is in eV and is calculated using PBE$+$SOC (the HSE06$+$SOC derived $E_{\rm g}$ for Bi$_2$Se$_3$ is 0.28\,eV). Our results are compared to available experimental and theoretical values in the literature. The van der Waals density functional with Cooper's exchange (vdW-DF$^{\rm C09}_{x}$) was used in Ref.\,\onlinecite{Luo2012}.}
 \begin{ruledtabular}
    \begin{tabular}[c]{ccccc}
    & & Bi$_2$Se$_3$ & Bi$_2$Te$_3$ & Sb$_2$Te$_3$\\
     \hline
      & {\it normal} & 4.092 & 4.349 & 4.195\\
      & {\it fcc} & 4.071 & 4.290 & 4.144\\
   $a$ & {\it top} & 4.057 & 4.282 & 4.142\\
      & Experiment & 4.138\footnotemark[1] & 4.383\footnotemark[1] & 4.250\footnotemark[1]\\
      & Theory ({\it normal}) & 4.125\footnotemark[2] & 4.360\footnotemark[2], 4.530\footnotemark[3] & 4.440\footnotemark[3]\\
      \\
      & {\it normal} & 28.91 & 31.09 & 30.81\\
      & {\it fcc} & 29.60 & 32.03 & 31.85\\
  $c$ & {\it top} & 32.15 & 34.74 &  34.53\\
      & Experiment & 28.64\footnotemark[1] & 30.49\footnotemark[1] & 30.35\footnotemark[1]\\
      & Theory ({\it normal}) & 28.76\footnotemark[2] & 30.17\footnotemark[2], 30.63\footnotemark[3] & 30.29\footnotemark[3]\\
      \\
            & {\it normal} & 0.36 & 0.15 & 0.13\\
        $E_{\rm g}$ & Experiment & 0.35\footnotemark[4] & 0.17\footnotemark[5],0.15\footnotemark[4], 0.13\footnotemark[6] & 0.30\footnotemark[4], 0.21\footnotemark[6]\\
      & Theory ({\it normal}) & 0.30\footnotemark[7], 0.31\footnotemark[8] & 0.08\footnotemark[8], 0.05\footnotemark[3] & 0.03\footnotemark[3]\\
      \\
          & {\it normal} & $-$4.30 & $-$3.82 & $-$3.17\\
    $E_{\rm coh}$  & {\it fcc} & $-$4.00 & $-$3.80 & $-$3.16\\
   & {\it top} & $-$3.98 & $-$3.76 &  $-$3.04\\
    \end{tabular}
    \end{ruledtabular}
\footnotetext{Reference\,\onlinecite{Wyckoff1964}}
\footnotetext{Reference\,\onlinecite{Luo2012}}
\footnotetext{Reference\,\onlinecite{Wang2007}}
\footnotetext{Reference\,\onlinecite{Black1957}}
\footnotetext{Reference\,\onlinecite{Chen2009}}
\footnotetext{Reference\,\onlinecite{Sehr1962}}
\footnotetext{Reference\,\onlinecite{Zhang2009}}
\footnotetext{Reference\,\onlinecite{Yazyev2012}}
\label{tab1}
\end{table*}	

\subsection{Ultrathin V$_2$VI$_3$ nanofilms}
\subsubsection{Energetics and thermodynamic stability of V$_2$VI$_3$ nanofilms}
\begin{figure}[tb!]
\center
\includegraphics[width=0.50\textwidth]{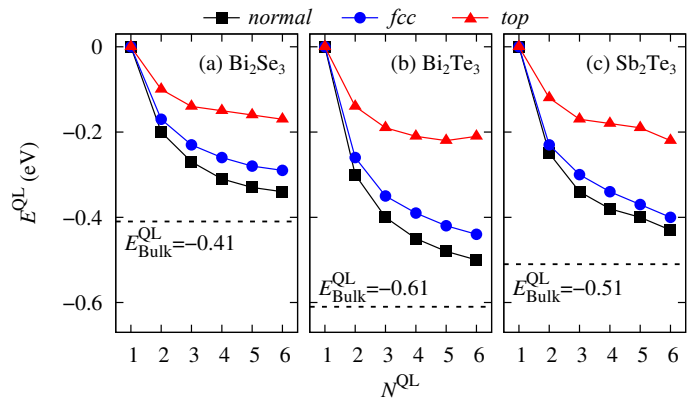}
\caption{(Color online) QL-QL interaction energy, $E^{\rm QL}$ of {\it normal}, {\it fcc}, and {\it top} polymorphic nanofilms as a function of increasing number of QLs, $N^{\rm QL}$ for (a) Bi$_2$Se$_3$, (b) Bi$_2$Te$_3$, and (c) Sb$_2$Te$_3$. As an example, the bulk-limit of this interaction energy, $E^{\rm QL}_{\rm bulk}$ is shown as the horizontal dotted line for each V$_2$VI$_3$ compound.}
\label{fig2}
\end{figure}

For the nanofilms of these V$_2$VI$_3$ materials, we perform PBE$+$D2($+$SOC) calculations to study their QL-QL interaction energy, $E^{\rm QL}$, as well as to understand their electronic band properties as a function of increasing number of QLs, $N^{\rm QL}$. Figure\,\ref{fig2} shows the variation of $E^{\rm QL}$ for {\it normal}, {\it fcc}, and {\it top} structured nanofilms as a function of $N^{\rm QL}$ for Bi$_2$Se$_3$ (Fig.\,\ref{fig2}a), Bi$_2$Te$_3$ (Fig.\,\ref{fig2}b), and Sb$_2$Te$_3$ (Fig.\,\ref{fig2}c), respectively. $E^{\rm QL}$ for all three stackings considered in this work are calculated to be negative, and converge to the bulk-like values ($E^{\rm QL}_{\rm bulk}$) with increasing $N^{\rm QL}$. Taking the bulk-stacked {\it normal} structure as an example, we find that $E^{\rm QL}_{\rm bulk}$ $= -0.41$, $-0.61$, and $-0.51$ eV for Bi$_2$Se$_3$, Bi$_2$Te$_3$, and Sb$_2$Te$_3$, respectively. The {\it top} structured nanofilms have the least favorable $E^{\rm QL}$, while the naturally forming {\it normal} structure yield the most favorable $E^{\rm QL}$, with almost similar values for {\it fcc} stacked nanofilms. These results are also consistent with the calculated cohesive energies of their bulk polymorphs, as illustrated above. Given that the calculated average difference in $E^{\rm QL}$ between the {\it normal} and {\it fcc} polymorphs is only about 0.06\,eV, it is not difficult to imagine metastable {\it fcc} forming from {\it normal} stacked films via a $martensitic$-like (i.e. diffusionless) planar displacement, especially when the film thickness is in the nanometer range. 

\begin{figure}[tb!]
\center
\includegraphics[width=0.40\textwidth]{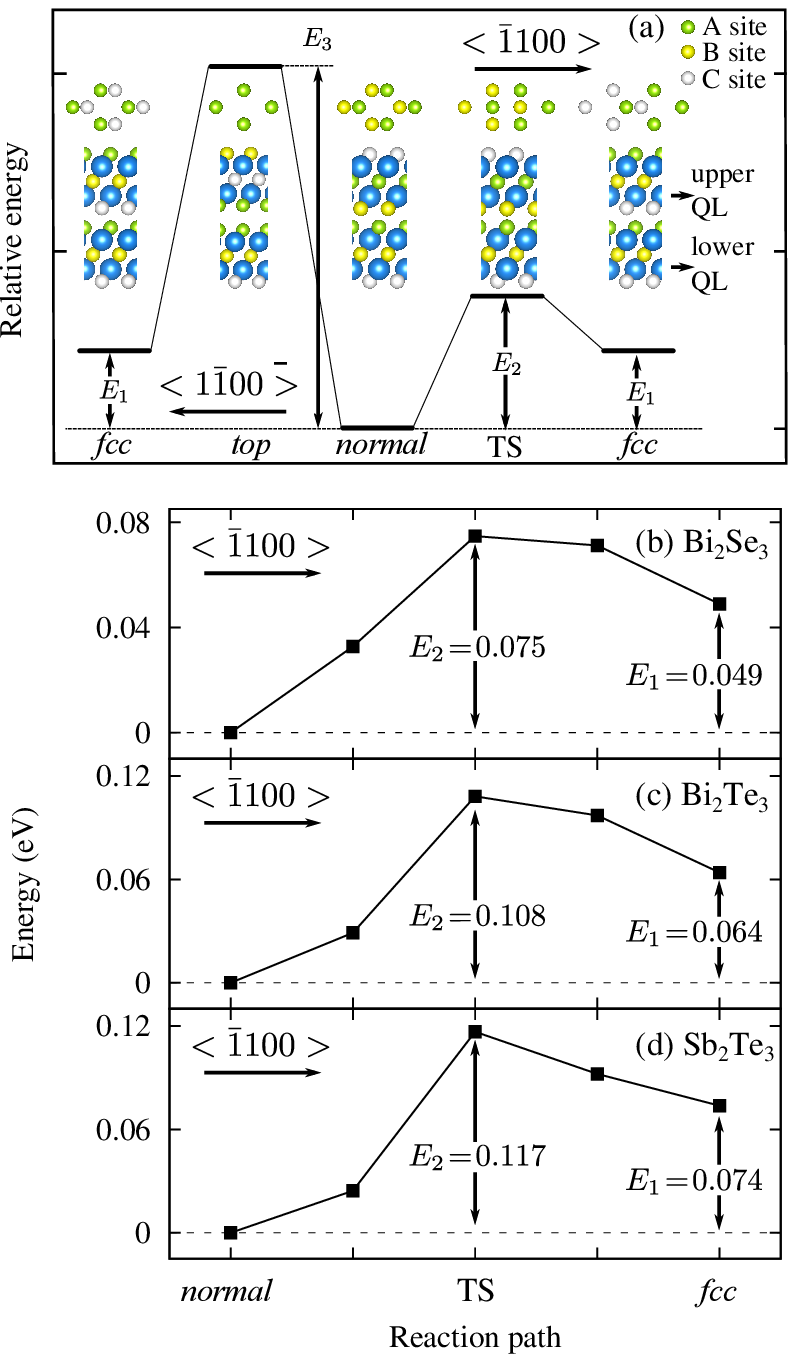}
\caption{(Color online) Energy profile for $slippage$: Relative energy differences between the {\it normal} and {\it fcc} structures for 2\,QL nanofilms for two different paths, namely along the $\langle\bar{1}100\rangle$ (energy $E_{2}$) or $\langle 1\bar{1}00\rangle$ (energy $E_{3}$) directions. Schematic energy profile along the $\langle\bar{1}100\rangle$ direction for: (b) Bi$_2$Se$_3$, (c) Bi$_2$Te$_3$, and (d) Sb$_2$Te$_3$. All energies are given with respect to the reference total energy of {\it normal} films.}
\label{fig3}
\end{figure}

To study the energetic profile of this possible planar displacement, we use the the simplest 2\,QL nanofilms for each V$_2$VI$_3$ to study planar displacement (here, we term this as $slippage$). Similar to the slippage in other 2D materials e.g. bilayer graphene and boron nitride,\cite{Lebedeva2011,Liu2003} the upper QL may slip along the $\langle\bar{1}100\rangle$ or $\langle 1\bar{1}00\rangle$ directions as shown in Fig.\,\ref{fig3}a. In these two directions, {\it normal}, {\it fcc}, and {\it top} structure will appear when the bottom Se atomic layer in the upper QL is located in B, C and A sites, respectively. Since each QL is one unit, movement appears only between adjoining QLs. To understand the barrier needed to undergo this slippage in the $\langle\bar{1}100\rangle$ direction, we have also considered possible transition states (TS) structures between the {\it normal} and the {\it fcc} structured nanofilm. For the other $\langle 1\bar{1}00\rangle$ direction, the pathway goes via the {\it top} polymorph. These mid-way configurations are consistent with the slippage in bilayer graphene and boron nitride.\cite{Sakamoto2010} 

To calculate these slippage energy barriers, the energies of $E_{1}\, (= E_{fcc} - E_{normal}),$ $E_{2}\, (= E_{\rm TS} - E_{normal})$ and $E_{3}\, (= E_{top} - E_{normal})$ have been defined in Fig.\,\ref{fig3}a, where $E_{normal}$, $E_{fcc}$, $E_{top}$, and $E_{\rm TS}$ denote the energies of {\it normal}, {\it fcc}, and {\it top} structure as well as the transition state, respectively. As shown in Figs.\,\ref{fig3}b to \ref{fig3}d, the energies of {\it top} with respect to {\it normal} structured films, $E_{1}$ is calculated to be 0.049, 0.064, and 0.074\,eV for Bi$_2$Se$_3$, Bi$_2$Te$_3$, and Sb$_2$Te$_3$, respectively. And based on our CI-NEB calculations, for the energy barrier in the $\langle\bar{1}100\rangle$ direction, $E_{2}$ is found to be 0.075, 0.108, and 0.117\,eV for Bi$_2$Se$_3$, Bi$_2$Te$_3$, and Sb$_2$Te$_3$, respectively, while for the $\langle 1\bar{1}00\rangle$ direction, $E_{3}$ is found to be 0.205, 0.254, and 0.295\,eV, correspondingly. We see that $E_{3}$ is at least two to three times larger than $E_{2}$ for all cases. Compared to the reported $E^{\rm QL}$ (19.3\,meV) and area-normalized $E_{2}$ (1.6\,meV/{\AA}$^2$) values for bilayer graphene,\cite{Lebedeva2011} we find rather similar orders of magnitude for the V$_2$VI$_3$ chalcogenide nanofilms: 13.8 and 4.8\,meV/{\AA}$^2$ for Bi$_2$Se$_3$ films, 18.3 and 5.5\,meV/{\AA}$^2$ for Bi$_2$Te$_3$ films, and 16.4 and 6.5\,meV/{\AA}$^2$ for Sb$_2$Te$_3$ films, respectively. This would then imply that the martensitic {\it normal}-to-{\it fcc} slippage would most probably occur for these V$_2$VI$_3$ chalcogenide nanofilms via the $\langle\bar{1}100\rangle$ rather than the $\langle 1\bar{1}00\rangle$ direction. 

\subsubsection{Electronic structure of V$_2$VI$_3$ nanofilms}
\begin{figure}[tb!]
\center
\includegraphics[width=0.45\textwidth]{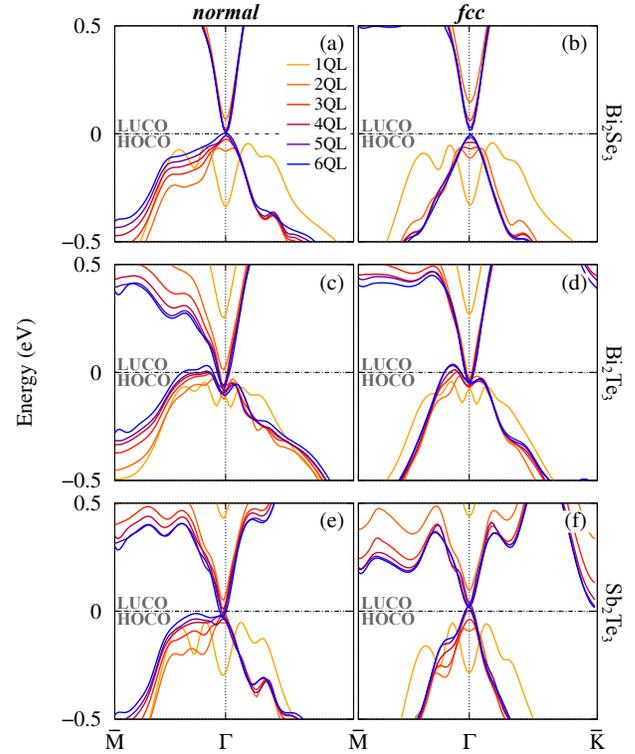}
\caption{(Color online) PBE$+$SOC DFT-derived QL-resolved electronic band structure of V$_2$VI$_3$ nanofilms (from 1 to 6\,QL): (a) {\it normal} structure of Bi$_2$Se$_3$, (b) {\it fcc} structure of Bi$_2$Se$_3$, (c) {\it normal} structure of Bi$_2$Te$_3$, (d) {\it fcc} structure of Bi$_2$Te$_3$, (e) {\it normal} structure of Sb$_2$Te$_3$, and (f) {\it fcc} structure of Sb$_2$Te$_3$. Only the band-edges due to the highest occupied crystal orbital (HOCO) and the lowest unoccupied crystal orbital (LUCO) are shown for clarity.}
\label{fig4}
\end{figure}

\begin{figure}[tb!]
\center
\includegraphics[width=0.50\textwidth]{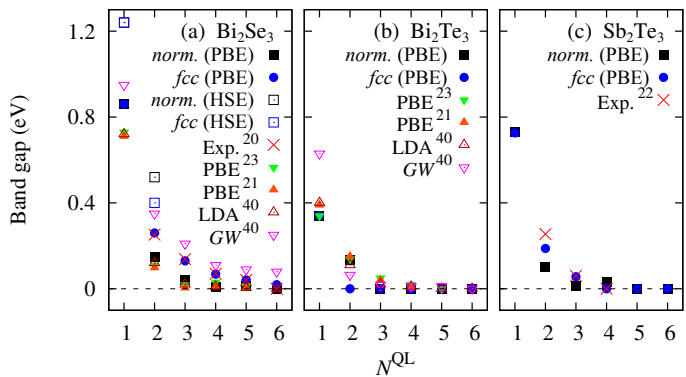}
\caption{(Color online) $\Gamma$-point band gap energies, $E_{\rm g}$ as a function of the number of QL, $N^{\rm QL}$: (a) Bi$_2$Se$_3$, (b) Bi$_2$Te$_3$, and (c) Sb$_2$Te$_3$. {\it normal} (PBE) and {\it fcc} (PBE) denote the PBE$+$SOC value, while {\it normal} (HSE) and {\it fcc} (HSE) denote that calculated by the HSE06$+$SOC hybrid functional, respectively. The absolute values for these $E_{\rm g}$ are listed in Table\,\ref{tabS1} of the Appendix section.}
\label{fig5}
\end{figure}

Turning to the electronic structure of the {\it normal} and {\it fcc} stacked nanofilms of V$_2$VI$_3$, we study both the electronic band dispersion, as well as the electron density plots (at the band-edges) to understand how their topological insulating properties might be influenced by both stacking and film thickness in these nanosystems. In Fig.\,\ref{fig4}, we calculate and plot the electronic band structures (using PBE$+$SOC), showing only the highest occupied crystal orbital (HOCO) and the lowest unoccupied crystal orbital (LUCO) for different numbers of QLs. The full band structures are shown in Fig.\,\ref{figS1} of the Appendix section, where the number of bands necessarily increases linearly with the thickness of the films. The overall band dispersion profile (especially near the $\Gamma$ point) of the {\it normal} structured films are found to be rather similar to those of the corresponding {\it fcc} polymorph, but the clear difference lies in the variation in the magnitude of $E_{\rm g}$. To cross-check our PBE+SOC results, we have also selectively calculated the electronic structure of 1 and 2\,QL {\it normal} and {\it fcc} stacking of Bi$_2$Se$_3$ using HSE06+SOC where the predicted $E_{\rm g}$ are plotted in Fig.\,\ref{fig5}.

For Bi$_2$Se$_3$, $E_{\rm g}$ decreases to zero when the $N^{\rm QL}$ of {\it normal} structured nanofilms increases to 6, while that of all {\it fcc}-like films remain greater than zero. The predicted $E_{\rm g}$ of Bi$_2$Se$_3$ {\it normal} structured films are in good agreement with other theoretical reports,\cite{Liu2010,Yazyev2010,Yazyev2012,Luo2012} but differ from the reported experimental data for Bi$_2$Se$_3$ films,\cite{Zhang2010} as shown in Fig.\,\ref{fig5}a. Interestingly, the predicted $E_{\rm g}$ of the {\it fcc} polymorphs, except for the 6\,QL nanofilm, seem to agree closely with the experimental $E_{\rm g}$ of Bi$_2$Se$_3$ films. Accordingly, the predicted $E_{\rm g}$ of Bi$_2$Se$_3$ {\it fcc} structured films are systematically larger than that of the Bi$_2$Se$_3$ {\it normal} stacked. Our HSE06+SOC derived $E_{\rm g}$ for 2\,QL {\it normal} and {\it fcc} structure show the same tendency (albeit having larger absolute values of $E_{\rm g}$, and poorer agreement with reported experimental values). The $E_{\rm g}$ of the {\it fcc} structured 6\,QL is close to zero (0.02\,eV), which may lead to a closure of the band gap when the $N^{\rm QL}$ increases continually. For Bi$_2$Te$_3$, as shown in Fig.\,\ref{fig5}b, the $E_{\rm g}$ of the {\it normal} polymorph decreases to zero when $N^{\rm QL}$ is increased to 3, again agreeing with other theoretical reports.\cite{Liu2010,Yazyev2010,Yazyev2012,Luo2012} In contrast to Bi$_2$Se$_3$, the $E_{\rm g}$ of the {\it fcc} structured films quickly reaches zero when $N^{\rm QL}$ reaches 2. For Sb$_2$Te$_3$, $E_{\rm g}$ of all {\it normal} structural films are greater than (or close to) zero for $N^{\rm QL}$ less than 5 while $E_{\rm g}$ decreases to zero when the $N^{\rm QL}$ of {\it fcc} stacked films increases to 4, cf. Fig.\,\ref{fig5}c. Again, the reported experimental data\cite{Jiang2012} seems to agree better with the predicted $E_{\rm g}$ of the {\it fcc} polymorph. 

Based on the above comparison, within the accuracy of PBE$+$SOC DFT, the experimentally measured values of $E_{\rm g}$ for these V$_2$VI$_3$ nanofilms seem to better match those of V$_2$VI$_3$ {\it fcc} structured nanofilms, rather than the assumed more stable {\it normal} stacking.  Although from our bulk calculations of Bi$_2$Se$_3$, we find that the PBE$+$SOC $E_{\rm g}$ value is found to agree better with available experimental and $GW$ values, we cannot rule out the intrinsic deficiencies of semi-local DFT which could underestimate the predicted $E_{\rm g}$ values for these polymorphic nanofilms. We also note that a self-consistent relativistic quasiparticle treatment may change the quantitative nature of our results, but the qualitative trends discussed here are expected to be reliable.

\begin{figure}[tb!]
\center
\includegraphics[width=0.40\textwidth]{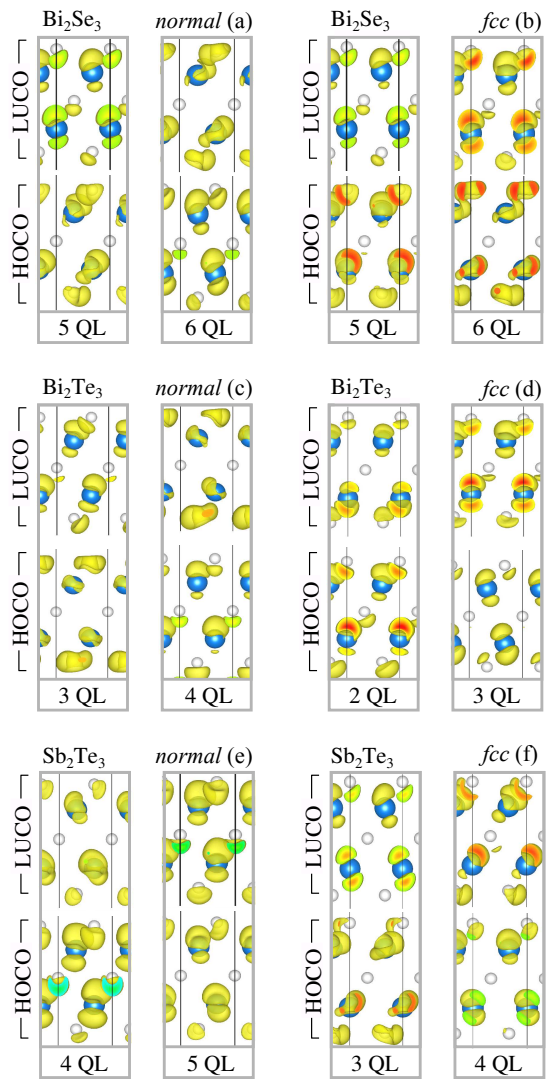}
\caption{(Color online) $\Gamma$-point electron density distributions (namely the HOCO and LUCO states): (a) 5 and 6\,QL {\it normal} structured films of Bi$_2$Se$_3$, (b) 5 and 6\,QL {\it fcc} structured films of Bi$_2$Se$_3$, (c) 3 and 4\,QL {\it normal} structured films of Bi$_2$Te$_3$, (d) 2 and 3\,QL {\it fcc} structured films of Bi$_2$Te$_3$, (e) 3 and 4\,QL {\it normal} structured films of Sb$_2$Te$_3$, and (f) 3 and 4\,QL {\it fcc} structured films of Sb$_2$Te$_3$.}
\label{fig6}
\end{figure}

Although vanishing values of $E_{\rm g}$ for some nanofilms are predicted, this is not sufficient information to conclude that these states are topologically protected. For systems with inversion symmetry, such as the V$_2$VI$_3$ compounds considered in this work, the Z$_2$ topological order can be determined by a parity analysis of the occupied states at time-reversal points in the Brillouin zone.\cite{Yazyev2012,Fu2007} As a first approximation to this full Z$_2$ topological analysis,\cite{Liu2011} we plot and study the $\Gamma$ point electron density distributions (namely the HOCO and LUCO states) for a selection of films in Fig.\,\ref{fig6}. 

Taking the {\it normal} stacked 5\,QL Bi$_2$Se$_3$ as a starting example, the HOCO state is more localized on the more electronegative Se anion, while the electron density of the LUCO is mainly concentrated on the Bi cation, which is typical of a normal insulator. However, upon increasing $N^{\rm QL}$ to 6, the orbital symmetry of the HOCO and LUCO of the {\it normal} stacked 6\,QL film inverts (as compared to that of 5\,QL; see Fig.\,\ref{fig6}a). The electron density plot of the 6\,QL {\it normal} structured films LUCO resembles that of the HOCO of the 5\,QL film, and vice versa. This inversion of the orbital character suggests a possible transition from a normal insulator (NI) to a TI, which has been observed in experimental measurements.\cite{Zhang2010} However, for the {\it fcc} stacked Bi$_2$Se$_3$ films, the calculated HOCO and LUCO plots for $N^{\rm QL}$ ranging from 1 to 6 show very similar orbital character, suggesting that no such NI to TI inversion has taken place. We illustrate this for the 5 and 6\,QL {\it fcc} structured films in Fig.\,\ref{fig6}b. For the Bi$_2$Te$_3$ nanofilms, an orbital parity inversion is observed for both stackings {\it normal} and {\it fcc}, taking place at the increase from 3\,QL to 4\,QL for the {\it normal} structure, cf. Fig.\,\ref{fig6}c, and for the {\it fcc}-like films the inversion is observed when $N^{\rm QL}$ is increased from 2 to 3. And likewise, based on the electron density distribution for Sb$_2$Te$_3$, the orbital parity is predicted to be inverted for both {\it normal} and {\it fcc} structure (i.e. 4 to 5\,QL for {\it normal} and 3 to 4\,QL for {\it fcc}). 1, the critical change from 3 to 4\,QL in {\it fcc} structured films matches the experimental results better.\cite{Jiang2012} The inversion at a lower number of QLs for the {\it fcc} stacking may be due to a reduced overlap of the surface states, since the {\it fcc} structured slabs are thicker than the {\it normal} stacked, cf. also Table\,\ref{tab1}.

Upon studying the band gap, $E_{\rm g}$, and orbital character inversion as a function of $N^{\rm QL}$ for the {\it normal} and {\it fcc} structure of these V$_2$VI$_3$ nanofilms, it seems to suggest that given the very low energy barriers for the {\it normal}-to-{\it fcc} slippage, metastable {\it fcc} structured nanofilms of these V$_2$VI$_3$ compounds could well form and offer different electronic band properties (e.g. critical $N^{\rm QL}$ for orbital parity inversion), especially for Bi$_2$Se$_3$ nanofilms. Thus, when these layered V$_2$VI$_3$ nanofilms are used in nano-devices, due to the weak interaction between these QLs, a mechanical slip could easily be experienced and thus the desired topological insulating character of these films could be intentionally exploited for new technologies.

\section{Conclusion and summary}

The energetics of stacking sequences and their effect on the transition from normal to topological insulating behavior in Bi$_2$Se$_3$, Bi$_2$Te$_3$ and Sb$_2$Te$_3$ nanofilms have been investigated via first-principles DFT calculations. We find that the overall band dispersion is relatively insensitive to the stacking sequence, while the magnitude of the band gap, and the critical thickness for a band edge parity inversion, is more sensitive. Relatively low energy barriers may allow for martensitic {\it normal}-to-{\it fcc} slippage, which are predicted to change the electronic structure and alter the topological behavior of these V$_2$VI$_3$ nanofilms. Thus, the effect of mechanical shear stress should be carefully considered in applications of topological insulators, e.g. the nanotribological conversion from TI to NI may be exploited in nano-circuit switches. In this context, a more detailed experimental analysis of the present and accessible stacking sequences is called for. 

\begin{acknowledgments}
The authors gratefully acknowledge support by the Global Frontier R\,\&\,D Program (2013M3A6B1078881) on Center for Global Frontier Hybrid Interface Materials (GFHIM) funded by the Korean Ministry of Science, ICT\,\&\,Future Planning, as well as the Australian Research Council (ARC). This work was also supported by the third Stage of Brain Korea 21 Plus Project Division of Creative Materials. Computational resources have been provided by the Australian National Computational Infrastructure (NCI) and by the KISTI supercomputing center (KSC-2013-C3-040). A. W. and B. Y. acknowledge support from the Royal Society University Research Fellowship scheme, and the European Research Council Advanced Grant (ERC 291472), respectively.
\end{acknowledgments}

\onecolumngrid
\newpage
\appendix*
\section{Supporting Information}
\begin{figure*}[h!]
\center
\includegraphics[width=0.60\textwidth]{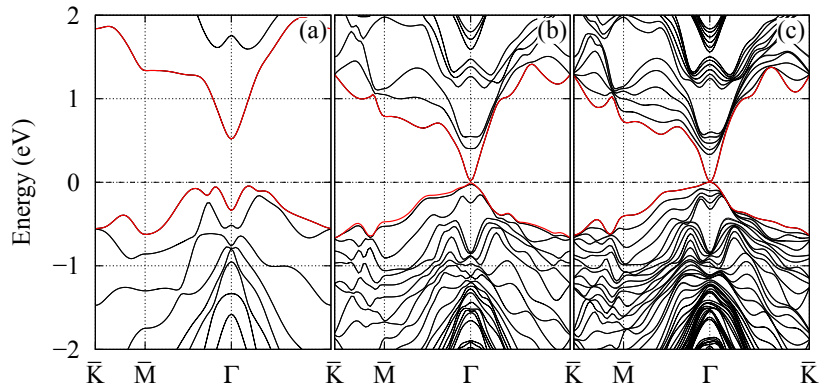}
\caption{(Color online) PBE$+$SOC DFT-derived electronic band structures of {\it normal} structured nanofilms of Bi$_2$Se$_3$: (a) $N^{\rm QL} = 1$, (b) $N^{\rm QL} = 3$, and (c) $N^{\rm QL} = 6$. The band-edges due to the highest occupied crystal orbital (HOCO) and the lowest unoccupied crystal orbital (LUCO) are shown in red lines.}
\label{figS1}
\end{figure*}

\begin{table*}[h!]
\centering
\caption{$\Gamma$-point band gap energies, $E_{\rm g}$ for {\it normal} and {\it fcc} structured nanofilms calculated using PBE+SOC (values in parenthesis by HSE06+SOC) as a function of the number of QL, $N^{\rm QL}$ for Bi$_2$Se$_3$, Bi$_2$Te$_3$, and Sb$_2$Te$_3$. These are compared to available experimental as well as theoretical values. All values are reported in eV.}
 \begin{ruledtabular}
    \begin{tabular}[c]{ccccccc}
    & $N^{\rm QL}$ & {\it normal} & {\it fcc} & Experiment & Theory\\
    \hline
    & 1 & 0.86 (1.19) & 0.86 (1.19)& & 0.95\footnotemark[2], 0.72\footnotemark[3], 0.73\footnotemark[3], 0.71\footnotemark[5] \\
    & 2 & 0.15 (0.36)& 0.26 (0.48)& 0.25\footnotemark[1] & 0.35\footnotemark[2], 0.12\footnotemark[3], 0.11\footnotemark[4], 0.098\footnotemark[5] \\
  Bi$_2$Se$_3$  & 3 & 0.04& 0.13 & 0.14\footnotemark[1] & 0.21\footnotemark[2], 0.02\footnotemark[3], 0.02\footnotemark[4], 0.004\footnotemark[5] \\
    & 4 & 0.01& 0.07 & 0.07\footnotemark[1] & 0.11\footnotemark[2], 0.01\footnotemark[3], 0.03\footnotemark[4], 0.01\footnotemark[5] \\
    & 5 & 0.01& 0.04 & 0.04\footnotemark[1] & 0.08\footnotemark[2], 0.00\footnotemark[3] \\
    & 6 & 0.00 & 0.02 & 0.00\footnotemark[1] &  \\
    \\ 
    & 1 & 0.34 & 0.34 & & 0.63\footnotemark[2], 0.40\footnotemark[3], 0.34\footnotemark[4], 0.39\footnotemark[5] \\
    & 2 & 0.14& 0.00 & & 0.07\footnotemark[2], 0.11\footnotemark[3], 0.14\footnotemark[4], 0.15\footnotemark[5] \\
  Bi$_2$Te$_3$  & 3 & 0.00 & 0.00 & &  0.02\footnotemark[2], 0.02\footnotemark[3], 0.05\footnotemark[4], 0.04\footnotemark[5] \\
    & 4 & 0.00 & 0.00 & &  0.01\footnotemark[2], 0.01\footnotemark[3], 0.02\footnotemark[4], 0.005\footnotemark[5] \\
    & 5 & 0.00 & 0.00 & &  0.01\footnotemark[2], 0.00\footnotemark[3], 0.00\footnotemark[4] \\
    & 6 & 0.00 & 0.00 & & 0.001\footnotemark[2], 0.00\footnotemark[3]\\
    \\
    & 1 & 0.73 & 0.73 & 0.67\footnotemark[6] & \\
    & 2 & 0.10 & 0.19 & 0.26\footnotemark[6] & \\
 Sb$_2$Te$_3$   & 3 & 0.01 & 0.06 & 0.06\footnotemark[6] & \\
    & 4 & 0.03 & 0.00 & 0.00\footnotemark[6] & \\
    & 5 & 0.00 & 0.00 & & \\
    & 6 & 0.00 & 0.00 & & \\
    \end{tabular}
   \end{ruledtabular}
\footnotetext{Experiment, Reference\,\onlinecite{Chen2011}}
\footnotetext{$G_0W_0$, Reference\,\onlinecite{Franchini2005}}
\footnotetext{LDA, Reference\,\onlinecite{Franchini2005}}
\footnotetext{PBE, Reference\,\onlinecite{Yazyev2010}}
\footnotetext{PBE, Reference\,\onlinecite{Liu2010}}
\footnotetext{Experiment, Reference\,\onlinecite{Li2010}}
\label{tabS1}
\end{table*}

\twocolumngrid	
\clearpage
\newpage


\begin{thebibliography}{50}%
\makeatletter
\providecommand \@ifxundefined [1]{%
 \@ifx{#1\undefined}
}%
\providecommand \@ifnum [1]{%
 \ifnum #1\expandafter \@firstoftwo
 \else \expandafter \@secondoftwo
 \fi
}%
\providecommand \@ifx [1]{%
 \ifx #1\expandafter \@firstoftwo
 \else \expandafter \@secondoftwo
 \fi
}%
\providecommand \natexlab [1]{#1}%
\providecommand \enquote  [1]{``#1''}%
\providecommand \bibnamefont  [1]{#1}%
\providecommand \bibfnamefont [1]{#1}%
\providecommand \citenamefont [1]{#1}%
\providecommand \href@noop [0]{\@secondoftwo}%
\providecommand \href [0]{\begingroup \@sanitize@url \@href}%
\providecommand \@href[1]{\@@startlink{#1}\@@href}%
\providecommand \@@href[1]{\endgroup#1\@@endlink}%
\providecommand \@sanitize@url [0]{\catcode `\\12\catcode `\$12\catcode
  `\&12\catcode `\#12\catcode `\^12\catcode `\_12\catcode `\%12\relax}%
\providecommand \@@startlink[1]{}%
\providecommand \@@endlink[0]{}%
\providecommand \url  [0]{\begingroup\@sanitize@url \@url }%
\providecommand \@url [1]{\endgroup\@href {#1}{\urlprefix }}%
\providecommand \urlprefix  [0]{URL }%
\providecommand \Eprint [0]{\href }%
\providecommand \doibase [0]{http://dx.doi.org/}%
\providecommand \selectlanguage [0]{\@gobble}%
\providecommand \bibinfo  [0]{\@secondoftwo}%
\providecommand \bibfield  [0]{\@secondoftwo}%
\providecommand \translation [1]{[#1]}%
\providecommand \BibitemOpen [0]{}%
\providecommand \bibitemStop [0]{}%
\providecommand \bibitemNoStop [0]{.\EOS\space}%
\providecommand \EOS [0]{\spacefactor3000\relax}%
\providecommand \BibitemShut  [1]{\csname bibitem#1\endcsname}%
\let\auto@bib@innerbib\@empty
\bibitem [{\citenamefont {Bernevig}\ \emph {et~al.}(2006)\citenamefont
  {Bernevig}, \citenamefont {Hughes},\ and\ \citenamefont
  {Zhang}}]{Bernevig2006}%
  \BibitemOpen
  \bibfield  {author} {\bibinfo {author} {\bibfnamefont {B.~A.}\ \bibnamefont
  {Bernevig}}, \bibinfo {author} {\bibfnamefont {T.~L.}\ \bibnamefont
  {Hughes}}, \ and\ \bibinfo {author} {\bibfnamefont {S.-C.}\ \bibnamefont
  {Zhang}},\ }\href@noop {} {\bibfield  {journal} {\bibinfo  {journal}
  {Science}\ }\textbf {\bibinfo {volume} {314}},\ \bibinfo {pages} {1757}
  (\bibinfo {year} {2006})}\BibitemShut {NoStop}%
\bibitem [{\citenamefont {Hasan}\ and\ \citenamefont {Kane}(2010)}]{Hasan2010}%
  \BibitemOpen
  \bibfield  {author} {\bibinfo {author} {\bibfnamefont {M.~Z.}\ \bibnamefont
  {Hasan}}\ and\ \bibinfo {author} {\bibfnamefont {C.~L.}\ \bibnamefont
  {Kane}},\ }\href@noop {} {\bibfield  {journal} {\bibinfo  {journal} {Rev.
  Mod. Phys.}\ }\textbf {\bibinfo {volume} {82}},\ \bibinfo {pages} {3045}
  (\bibinfo {year} {2010})}\BibitemShut {NoStop}%
\bibitem [{\citenamefont {Moore}(2010)}]{Moore2010}%
  \BibitemOpen
  \bibfield  {author} {\bibinfo {author} {\bibfnamefont {J.~E.}\ \bibnamefont
  {Moore}},\ }\href@noop {} {\bibfield  {journal} {\bibinfo  {journal}
  {Nature}\ }\textbf {\bibinfo {volume} {464}},\ \bibinfo {pages} {194}
  (\bibinfo {year} {2010})}\BibitemShut {NoStop}%
\bibitem [{\citenamefont {Qi}\ and\ \citenamefont {Zhang}(2011)}]{Qi2011}%
  \BibitemOpen
  \bibfield  {author} {\bibinfo {author} {\bibfnamefont {X.-L.}\ \bibnamefont
  {Qi}}\ and\ \bibinfo {author} {\bibfnamefont {S.-C.}\ \bibnamefont {Zhang}},\
  }\href@noop {} {\bibfield  {journal} {\bibinfo  {journal} {Rev. Mod. Phys.}\
  }\textbf {\bibinfo {volume} {83}},\ \bibinfo {pages} {1057} (\bibinfo {year}
  {2011})}\BibitemShut {NoStop}%
\bibitem [{\citenamefont {Fu}\ and\ \citenamefont {Kane}(2007)}]{Fu2007}%
  \BibitemOpen
  \bibfield  {author} {\bibinfo {author} {\bibfnamefont {L.}~\bibnamefont
  {Fu}}\ and\ \bibinfo {author} {\bibfnamefont {C.~L.}\ \bibnamefont {Kane}},\
  }\href@noop {} {\bibfield  {journal} {\bibinfo  {journal} {Phys. Rev. B}\
  }\textbf {\bibinfo {volume} {76}},\ \bibinfo {pages} {045302} (\bibinfo
  {year} {2007})}\BibitemShut {NoStop}%
\bibitem [{\citenamefont {Hsieh}\ \emph {et~al.}(2008)\citenamefont {Hsieh},
  \citenamefont {Qian}, \citenamefont {Wray}, \citenamefont {Xia},
  \citenamefont {Hor}, \citenamefont {Cava},\ and\ \citenamefont
  {Hasan}}]{Hsieh2008}%
  \BibitemOpen
  \bibfield  {author} {\bibinfo {author} {\bibfnamefont {D.}~\bibnamefont
  {Hsieh}}, \bibinfo {author} {\bibfnamefont {D.}~\bibnamefont {Qian}},
  \bibinfo {author} {\bibfnamefont {L.}~\bibnamefont {Wray}}, \bibinfo {author}
  {\bibfnamefont {Y.}~\bibnamefont {Xia}}, \bibinfo {author} {\bibfnamefont
  {Y.~S.}\ \bibnamefont {Hor}}, \bibinfo {author} {\bibfnamefont {R.~J.}\
  \bibnamefont {Cava}}, \ and\ \bibinfo {author} {\bibfnamefont {M.~Z.}\
  \bibnamefont {Hasan}},\ }\href@noop {} {\bibfield  {journal} {\bibinfo
  {journal} {Nature}\ }\textbf {\bibinfo {volume} {452}} (\bibinfo {year}
  {2008})}\BibitemShut {NoStop}%
\bibitem [{\citenamefont {Yan}\ and\ \citenamefont {Zhang}(2012)}]{Yan2012}%
  \BibitemOpen
  \bibfield  {author} {\bibinfo {author} {\bibfnamefont {B.}~\bibnamefont
  {Yan}}\ and\ \bibinfo {author} {\bibfnamefont {S.-C.}\ \bibnamefont
  {Zhang}},\ }\href@noop {} {\bibfield  {journal} {\bibinfo  {journal} {Rep.
  Prog. Phys.}\ }\textbf {\bibinfo {volume} {75}},\ \bibinfo {pages} {096501}
  (\bibinfo {year} {2012})}\BibitemShut {NoStop}%
\bibitem [{\citenamefont {Zhang}\ \emph {et~al.}(2009)\citenamefont {Zhang},
  \citenamefont {Liu}, \citenamefont {Qi}, \citenamefont {Dai}, \citenamefont
  {Fang},\ and\ \citenamefont {Zhang}}]{Zhang2009}%
  \BibitemOpen
  \bibfield  {author} {\bibinfo {author} {\bibfnamefont {H.}~\bibnamefont
  {Zhang}}, \bibinfo {author} {\bibfnamefont {C.-X.}\ \bibnamefont {Liu}},
  \bibinfo {author} {\bibfnamefont {X.-L.}\ \bibnamefont {Qi}}, \bibinfo
  {author} {\bibfnamefont {X.}~\bibnamefont {Dai}}, \bibinfo {author}
  {\bibfnamefont {Z.}~\bibnamefont {Fang}}, \ and\ \bibinfo {author}
  {\bibfnamefont {S.-C.}\ \bibnamefont {Zhang}},\ }\href@noop {} {\bibfield
  {journal} {\bibinfo  {journal} {Nat. Phys.}\ }\textbf {\bibinfo {volume}
  {5}},\ \bibinfo {pages} {438} (\bibinfo {year} {2009})}\BibitemShut {NoStop}%
\bibitem [{\citenamefont {Wang}\ \emph {et~al.}(2011)\citenamefont {Wang},
  \citenamefont {Zhu}, \citenamefont {Sun}, \citenamefont {Li}, \citenamefont
  {Zhang}, \citenamefont {Wen}, \citenamefont {Chen}, \citenamefont {He},
  \citenamefont {Wang}, \citenamefont {Ma}, \citenamefont {Jia}, \citenamefont
  {Zhang},\ and\ \citenamefont {Xue}}]{Wang2011}%
  \BibitemOpen
  \bibfield  {author} {\bibinfo {author} {\bibfnamefont {G.}~\bibnamefont
  {Wang}}, \bibinfo {author} {\bibfnamefont {X.-G.}\ \bibnamefont {Zhu}},
  \bibinfo {author} {\bibfnamefont {Y.-Y.}\ \bibnamefont {Sun}}, \bibinfo
  {author} {\bibfnamefont {Y.-Y.}\ \bibnamefont {Li}}, \bibinfo {author}
  {\bibfnamefont {T.}~\bibnamefont {Zhang}}, \bibinfo {author} {\bibfnamefont
  {J.}~\bibnamefont {Wen}}, \bibinfo {author} {\bibfnamefont {X.}~\bibnamefont
  {Chen}}, \bibinfo {author} {\bibfnamefont {K.}~\bibnamefont {He}}, \bibinfo
  {author} {\bibfnamefont {L.-L.}\ \bibnamefont {Wang}}, \bibinfo {author}
  {\bibfnamefont {X.-C.}\ \bibnamefont {Ma}}, \bibinfo {author} {\bibfnamefont
  {J.-F.}\ \bibnamefont {Jia}}, \bibinfo {author} {\bibfnamefont {S.~B.}\
  \bibnamefont {Zhang}}, \ and\ \bibinfo {author} {\bibfnamefont {Q.-K.}\
  \bibnamefont {Xue}},\ }\href@noop {} {\bibfield  {journal} {\bibinfo
  {journal} {Adv. Mater.}\ }\textbf {\bibinfo {volume} {23}},\ \bibinfo {pages}
  {2929} (\bibinfo {year} {2011})}\BibitemShut {NoStop}%
\bibitem [{\citenamefont {Chen}\ \emph
  {et~al.}(2011{\natexlab{a}})\citenamefont {Chen}, \citenamefont {Ma},
  \citenamefont {He}, \citenamefont {Jia},\ and\ \citenamefont
  {Xue}}]{Chen2011}%
  \BibitemOpen
  \bibfield  {author} {\bibinfo {author} {\bibfnamefont {X.}~\bibnamefont
  {Chen}}, \bibinfo {author} {\bibfnamefont {X.-C.}\ \bibnamefont {Ma}},
  \bibinfo {author} {\bibfnamefont {K.}~\bibnamefont {He}}, \bibinfo {author}
  {\bibfnamefont {J.-F.}\ \bibnamefont {Jia}}, \ and\ \bibinfo {author}
  {\bibfnamefont {Q.-K.}\ \bibnamefont {Xue}},\ }\href@noop {} {\bibfield
  {journal} {\bibinfo  {journal} {Adv. Mater.}\ }\textbf {\bibinfo {volume}
  {23}},\ \bibinfo {pages} {1162} (\bibinfo {year}
  {2011}{\natexlab{a}})}\BibitemShut {NoStop}%
\bibitem [{\citenamefont {Scanlon}\ \emph {et~al.}(2012)\citenamefont
  {Scanlon}, \citenamefont {King}, \citenamefont {Singh}, \citenamefont {{de la
  Torre}}, \citenamefont {{McKeown Walker}}, \citenamefont {Balakrishnan},
  \citenamefont {Baumberger},\ and\ \citenamefont {Catlow}}]{Scanlon2012}%
  \BibitemOpen
  \bibfield  {author} {\bibinfo {author} {\bibfnamefont {D.~O.}\ \bibnamefont
  {Scanlon}}, \bibinfo {author} {\bibfnamefont {P.~D.~C.}\ \bibnamefont
  {King}}, \bibinfo {author} {\bibfnamefont {R.~P.}\ \bibnamefont {Singh}},
  \bibinfo {author} {\bibfnamefont {A.}~\bibnamefont {{de la Torre}}}, \bibinfo
  {author} {\bibfnamefont {S.}~\bibnamefont {{McKeown Walker}}}, \bibinfo
  {author} {\bibfnamefont {G.}~\bibnamefont {Balakrishnan}}, \bibinfo {author}
  {\bibfnamefont {F.}~\bibnamefont {Baumberger}}, \ and\ \bibinfo {author}
  {\bibfnamefont {C.~R.~A.}\ \bibnamefont {Catlow}},\ }\href@noop {} {\bibfield
   {journal} {\bibinfo  {journal} {Adv. Mater.}\ }\textbf {\bibinfo {volume}
  {24}},\ \bibinfo {pages} {2154} (\bibinfo {year} {2012})}\BibitemShut
  {NoStop}%
\bibitem [{\citenamefont {Chadov}\ \emph {et~al.}(2010)\citenamefont {Chadov},
  \citenamefont {Qi}, \citenamefont {K\"{u}bler}, \citenamefont {Fecher},\ and\
  \citenamefont {Felser}}]{Chadov2010}%
  \BibitemOpen
  \bibfield  {author} {\bibinfo {author} {\bibfnamefont {S.}~\bibnamefont
  {Chadov}}, \bibinfo {author} {\bibfnamefont {X.}~\bibnamefont {Qi}}, \bibinfo
  {author} {\bibfnamefont {J.}~\bibnamefont {K\"{u}bler}}, \bibinfo {author}
  {\bibfnamefont {G.~H.}\ \bibnamefont {Fecher}}, \ and\ \bibinfo {author}
  {\bibfnamefont {C.}~\bibnamefont {Felser}},\ }\href@noop {} {\bibfield
  {journal} {\bibinfo  {journal} {Nat. Mater.}\ }\textbf {\bibinfo {volume}
  {9}},\ \bibinfo {pages} {541} (\bibinfo {year} {2010})}\BibitemShut {NoStop}%
\bibitem [{\citenamefont {Lin}\ \emph {et~al.}(2010)\citenamefont {Lin},
  \citenamefont {Wray}, \citenamefont {Xia}, \citenamefont {Xu}, \citenamefont
  {Jia}, \citenamefont {Cava}, \citenamefont {Bansil},\ and\ \citenamefont
  {Hasan}}]{Lin2010}%
  \BibitemOpen
  \bibfield  {author} {\bibinfo {author} {\bibfnamefont {H.}~\bibnamefont
  {Lin}}, \bibinfo {author} {\bibfnamefont {L.~A.}\ \bibnamefont {Wray}},
  \bibinfo {author} {\bibfnamefont {Y.}~\bibnamefont {Xia}}, \bibinfo {author}
  {\bibfnamefont {S.}~\bibnamefont {Xu}}, \bibinfo {author} {\bibfnamefont
  {S.}~\bibnamefont {Jia}}, \bibinfo {author} {\bibfnamefont {R.~J.}\
  \bibnamefont {Cava}}, \bibinfo {author} {\bibfnamefont {A.}~\bibnamefont
  {Bansil}}, \ and\ \bibinfo {author} {\bibfnamefont {M.~Z.}\ \bibnamefont
  {Hasan}},\ }\href@noop {} {\bibfield  {journal} {\bibinfo  {journal} {Nat.
  Mater.}\ }\textbf {\bibinfo {volume} {9}},\ \bibinfo {pages} {546} (\bibinfo
  {year} {2010})}\BibitemShut {NoStop}%
\bibitem [{\citenamefont {Li}\ \emph {et~al.}(2011)\citenamefont {Li},
  \citenamefont {Lian},\ and\ \citenamefont {Jiang}}]{Li2011}%
  \BibitemOpen
  \bibfield  {author} {\bibinfo {author} {\bibfnamefont {C.}~\bibnamefont
  {Li}}, \bibinfo {author} {\bibfnamefont {J.~S.}\ \bibnamefont {Lian}}, \ and\
  \bibinfo {author} {\bibfnamefont {Q.}~\bibnamefont {Jiang}},\ }\href@noop {}
  {\bibfield  {journal} {\bibinfo  {journal} {Phys. Rev. B}\ }\textbf {\bibinfo
  {volume} {83}},\ \bibinfo {pages} {235125} (\bibinfo {year}
  {2011})}\BibitemShut {NoStop}%
\bibitem [{\citenamefont {Guo}\ and\ \citenamefont {Franz}(2009)}]{Guo2009}%
  \BibitemOpen
  \bibfield  {author} {\bibinfo {author} {\bibfnamefont {H.-M.}\ \bibnamefont
  {Guo}}\ and\ \bibinfo {author} {\bibfnamefont {M.}~\bibnamefont {Franz}},\
  }\href@noop {} {\bibfield  {journal} {\bibinfo  {journal} {Phys. Rev. Lett.}\
  }\textbf {\bibinfo {volume} {103}},\ \bibinfo {pages} {206805} (\bibinfo
  {year} {2009})}\BibitemShut {NoStop}%
\bibitem [{\citenamefont {Dzero}\ \emph {et~al.}(2010)\citenamefont {Dzero},
  \citenamefont {Sun}, \citenamefont {Galitski},\ and\ \citenamefont
  {Coleman}}]{Dzero2010}%
  \BibitemOpen
  \bibfield  {author} {\bibinfo {author} {\bibfnamefont {M.}~\bibnamefont
  {Dzero}}, \bibinfo {author} {\bibfnamefont {K.}~\bibnamefont {Sun}}, \bibinfo
  {author} {\bibfnamefont {V.}~\bibnamefont {Galitski}}, \ and\ \bibinfo
  {author} {\bibfnamefont {P.}~\bibnamefont {Coleman}},\ }\href@noop {}
  {\bibfield  {journal} {\bibinfo  {journal} {Phys. Rev. Lett.}\ }\textbf
  {\bibinfo {volume} {104}},\ \bibinfo {pages} {106408} (\bibinfo {year}
  {2010})}\BibitemShut {NoStop}%
\bibitem [{\citenamefont {Eremeev}\ \emph {et~al.}(2011)\citenamefont
  {Eremeev}, \citenamefont {Bihlmayer}, \citenamefont {Vergniory},
  \citenamefont {Koroteev}, \citenamefont {Menshchikova}, \citenamefont {Henk},
  \citenamefont {Ernst},\ and\ \citenamefont {Chulkov}}]{Eremeev2011}%
  \BibitemOpen
  \bibfield  {author} {\bibinfo {author} {\bibfnamefont {S.~V.}\ \bibnamefont
  {Eremeev}}, \bibinfo {author} {\bibfnamefont {G.}~\bibnamefont {Bihlmayer}},
  \bibinfo {author} {\bibfnamefont {M.}~\bibnamefont {Vergniory}}, \bibinfo
  {author} {\bibfnamefont {Y.~M.}\ \bibnamefont {Koroteev}}, \bibinfo {author}
  {\bibfnamefont {T.~V.}\ \bibnamefont {Menshchikova}}, \bibinfo {author}
  {\bibfnamefont {J.}~\bibnamefont {Henk}}, \bibinfo {author} {\bibfnamefont
  {A.}~\bibnamefont {Ernst}}, \ and\ \bibinfo {author} {\bibfnamefont {E.~V.}\
  \bibnamefont {Chulkov}},\ }\href@noop {} {\bibfield  {journal} {\bibinfo
  {journal} {Phys. Rev. B}\ }\textbf {\bibinfo {volume} {83}},\ \bibinfo
  {pages} {205129} (\bibinfo {year} {2011})}\BibitemShut {NoStop}%
\bibitem [{\citenamefont {Chen}\ \emph
  {et~al.}(2011{\natexlab{b}})\citenamefont {Chen}, \citenamefont {Zhu},
  \citenamefont {Xiao},\ and\ \citenamefont {Zhang}}]{Chen2011a}%
  \BibitemOpen
  \bibfield  {author} {\bibinfo {author} {\bibfnamefont {H.}~\bibnamefont
  {Chen}}, \bibinfo {author} {\bibfnamefont {W.}~\bibnamefont {Zhu}}, \bibinfo
  {author} {\bibfnamefont {D.}~\bibnamefont {Xiao}}, \ and\ \bibinfo {author}
  {\bibfnamefont {Z.}~\bibnamefont {Zhang}},\ }\href@noop {} {\bibfield
  {journal} {\bibinfo  {journal} {Phys. Rev. Lett.}\ }\textbf {\bibinfo
  {volume} {107}},\ \bibinfo {pages} {056804} (\bibinfo {year}
  {2011}{\natexlab{b}})}\BibitemShut {NoStop}%
\bibitem [{\citenamefont {Kong}\ and\ \citenamefont {Cui}(2011)}]{Kong2011}%
  \BibitemOpen
  \bibfield  {author} {\bibinfo {author} {\bibfnamefont {D.}~\bibnamefont
  {Kong}}\ and\ \bibinfo {author} {\bibfnamefont {Y.}~\bibnamefont {Cui}},\
  }\href@noop {} {\bibfield  {journal} {\bibinfo  {journal} {Nat. Chem.}\
  }\textbf {\bibinfo {volume} {3}},\ \bibinfo {pages} {845} (\bibinfo {year}
  {2011})}\BibitemShut {NoStop}%
\bibitem [{\citenamefont {Li}\ \emph {et~al.}(2010)\citenamefont {Li},
  \citenamefont {Wang}, \citenamefont {Zhu}, \citenamefont {Liu}, \citenamefont
  {Ye}, \citenamefont {Chen}, \citenamefont {Wang}, \citenamefont {He},
  \citenamefont {Wang}, \citenamefont {Ma}, \citenamefont {Zhang},
  \citenamefont {Dai}, \citenamefont {Fang}, \citenamefont {Xie}, \citenamefont
  {Liu}, \citenamefont {Qi}, \citenamefont {Jia}, \citenamefont {Zhang},\ and\
  \citenamefont {Xue}}]{Li2010}%
  \BibitemOpen
  \bibfield  {author} {\bibinfo {author} {\bibfnamefont {Y.-Y.}\ \bibnamefont
  {Li}}, \bibinfo {author} {\bibfnamefont {G.}~\bibnamefont {Wang}}, \bibinfo
  {author} {\bibfnamefont {X.-G.}\ \bibnamefont {Zhu}}, \bibinfo {author}
  {\bibfnamefont {M.-H.}\ \bibnamefont {Liu}}, \bibinfo {author} {\bibfnamefont
  {C.}~\bibnamefont {Ye}}, \bibinfo {author} {\bibfnamefont {X.}~\bibnamefont
  {Chen}}, \bibinfo {author} {\bibfnamefont {Y.-Y.}\ \bibnamefont {Wang}},
  \bibinfo {author} {\bibfnamefont {K.}~\bibnamefont {He}}, \bibinfo {author}
  {\bibfnamefont {L.-L.}\ \bibnamefont {Wang}}, \bibinfo {author}
  {\bibfnamefont {X.-C.}\ \bibnamefont {Ma}}, \bibinfo {author} {\bibfnamefont
  {H.-J.}\ \bibnamefont {Zhang}}, \bibinfo {author} {\bibfnamefont
  {X.}~\bibnamefont {Dai}}, \bibinfo {author} {\bibfnamefont {Z.}~\bibnamefont
  {Fang}}, \bibinfo {author} {\bibfnamefont {X.-C.}\ \bibnamefont {Xie}},
  \bibinfo {author} {\bibfnamefont {Y.}~\bibnamefont {Liu}}, \bibinfo {author}
  {\bibfnamefont {X.-L.}\ \bibnamefont {Qi}}, \bibinfo {author} {\bibfnamefont
  {J.-F.}\ \bibnamefont {Jia}}, \bibinfo {author} {\bibfnamefont {S.-C.}\
  \bibnamefont {Zhang}}, \ and\ \bibinfo {author} {\bibfnamefont {Q.-K.}\
  \bibnamefont {Xue}},\ }\href@noop {} {\bibfield  {journal} {\bibinfo
  {journal} {Adv. Mater.}\ }\textbf {\bibinfo {volume} {22}},\ \bibinfo {pages}
  {4002} (\bibinfo {year} {2010})}\BibitemShut {NoStop}%
\bibitem [{\citenamefont {Zhang}\ \emph {et~al.}(2010)\citenamefont {Zhang},
  \citenamefont {He}, \citenamefont {Chang}, \citenamefont {Song},
  \citenamefont {Wang}, \citenamefont {Chen}, \citenamefont {Jia},
  \citenamefont {Fang}, \citenamefont {Dai}, \citenamefont {Shan},
  \citenamefont {Shen}, \citenamefont {Niu}, \citenamefont {Qi}, \citenamefont
  {Zhang}, \citenamefont {Ma},\ and\ \citenamefont {Xue}}]{Zhang2010}%
  \BibitemOpen
  \bibfield  {author} {\bibinfo {author} {\bibfnamefont {Y.}~\bibnamefont
  {Zhang}}, \bibinfo {author} {\bibfnamefont {K.}~\bibnamefont {He}}, \bibinfo
  {author} {\bibfnamefont {C.-Z.}\ \bibnamefont {Chang}}, \bibinfo {author}
  {\bibfnamefont {C.-L.}\ \bibnamefont {Song}}, \bibinfo {author}
  {\bibfnamefont {L.-L.}\ \bibnamefont {Wang}}, \bibinfo {author}
  {\bibfnamefont {X.}~\bibnamefont {Chen}}, \bibinfo {author} {\bibfnamefont
  {J.-F.}\ \bibnamefont {Jia}}, \bibinfo {author} {\bibfnamefont
  {Z.}~\bibnamefont {Fang}}, \bibinfo {author} {\bibfnamefont {X.}~\bibnamefont
  {Dai}}, \bibinfo {author} {\bibfnamefont {W.-Y.}\ \bibnamefont {Shan}},
  \bibinfo {author} {\bibfnamefont {S.-Q.}\ \bibnamefont {Shen}}, \bibinfo
  {author} {\bibfnamefont {Q.}~\bibnamefont {Niu}}, \bibinfo {author}
  {\bibfnamefont {X.-L.}\ \bibnamefont {Qi}}, \bibinfo {author} {\bibfnamefont
  {S.-C.}\ \bibnamefont {Zhang}}, \bibinfo {author} {\bibfnamefont {X.-C.}\
  \bibnamefont {Ma}}, \ and\ \bibinfo {author} {\bibfnamefont {Q.-K.}\
  \bibnamefont {Xue}},\ }\href@noop {} {\bibfield  {journal} {\bibinfo
  {journal} {Nat. Phys.}\ }\textbf {\bibinfo {volume} {6}},\ \bibinfo {pages}
  {584} (\bibinfo {year} {2010})}\BibitemShut {NoStop}%
\bibitem [{\citenamefont {Liu}\ \emph {et~al.}(2010)\citenamefont {Liu},
  \citenamefont {Zhang}, \citenamefont {Yan}, \citenamefont {Qi}, \citenamefont
  {Frauenheim}, \citenamefont {Dai}, \citenamefont {Fang},\ and\ \citenamefont
  {Zhang}}]{Liu2010}%
  \BibitemOpen
  \bibfield  {author} {\bibinfo {author} {\bibfnamefont {C.-X.}\ \bibnamefont
  {Liu}}, \bibinfo {author} {\bibfnamefont {H.}~\bibnamefont {Zhang}}, \bibinfo
  {author} {\bibfnamefont {B.}~\bibnamefont {Yan}}, \bibinfo {author}
  {\bibfnamefont {X.-L.}\ \bibnamefont {Qi}}, \bibinfo {author} {\bibfnamefont
  {T.}~\bibnamefont {Frauenheim}}, \bibinfo {author} {\bibfnamefont
  {X.}~\bibnamefont {Dai}}, \bibinfo {author} {\bibfnamefont {Z.}~\bibnamefont
  {Fang}}, \ and\ \bibinfo {author} {\bibfnamefont {S.-C.}\ \bibnamefont
  {Zhang}},\ }\href@noop {} {\bibfield  {journal} {\bibinfo  {journal} {Phys.
  Rev. B}\ }\textbf {\bibinfo {volume} {81}},\ \bibinfo {pages} {041307}
  (\bibinfo {year} {2010})}\BibitemShut {NoStop}%
\bibitem [{\citenamefont {Jiang}\ \emph {et~al.}(2012)\citenamefont {Jiang},
  \citenamefont {Wang}, \citenamefont {Chen}, \citenamefont {Li}, \citenamefont
  {Song}, \citenamefont {He}, \citenamefont {Wang}, \citenamefont {Chen},
  \citenamefont {Ma},\ and\ \citenamefont {Xue}}]{Jiang2012}%
  \BibitemOpen
  \bibfield  {author} {\bibinfo {author} {\bibfnamefont {Y.}~\bibnamefont
  {Jiang}}, \bibinfo {author} {\bibfnamefont {Y.}~\bibnamefont {Wang}},
  \bibinfo {author} {\bibfnamefont {M.}~\bibnamefont {Chen}}, \bibinfo {author}
  {\bibfnamefont {Z.}~\bibnamefont {Li}}, \bibinfo {author} {\bibfnamefont
  {C.}~\bibnamefont {Song}}, \bibinfo {author} {\bibfnamefont {K.}~\bibnamefont
  {He}}, \bibinfo {author} {\bibfnamefont {L.}~\bibnamefont {Wang}}, \bibinfo
  {author} {\bibfnamefont {X.}~\bibnamefont {Chen}}, \bibinfo {author}
  {\bibfnamefont {X.}~\bibnamefont {Ma}}, \ and\ \bibinfo {author}
  {\bibfnamefont {Q.-K.}\ \bibnamefont {Xue}},\ }\href@noop {} {\bibfield
  {journal} {\bibinfo  {journal} {Phys. Rev. Lett.}\ }\textbf {\bibinfo
  {volume} {108}},\ \bibinfo {pages} {016401} (\bibinfo {year}
  {2012})}\BibitemShut {NoStop}%
\bibitem [{\citenamefont {Yazyev}\ \emph {et~al.}(2010)\citenamefont {Yazyev},
  \citenamefont {Moore},\ and\ \citenamefont {Louie}}]{Yazyev2010}%
  \BibitemOpen
  \bibfield  {author} {\bibinfo {author} {\bibfnamefont {O.~V.}\ \bibnamefont
  {Yazyev}}, \bibinfo {author} {\bibfnamefont {J.~E.}\ \bibnamefont {Moore}}, \
  and\ \bibinfo {author} {\bibfnamefont {S.~G.}\ \bibnamefont {Louie}},\
  }\href@noop {} {\bibfield  {journal} {\bibinfo  {journal} {Phys. Rev. Lett.}\
  }\textbf {\bibinfo {volume} {105}},\ \bibinfo {pages} {266806} (\bibinfo
  {year} {2010})}\BibitemShut {NoStop}%
\bibitem [{\citenamefont {Park}\ \emph {et~al.}(2010)\citenamefont {Park},
  \citenamefont {Heremans}, \citenamefont {Scarola},\ and\ \citenamefont
  {Minic}}]{Park2010}%
  \BibitemOpen
  \bibfield  {author} {\bibinfo {author} {\bibfnamefont {K.}~\bibnamefont
  {Park}}, \bibinfo {author} {\bibfnamefont {J.~J.}\ \bibnamefont {Heremans}},
  \bibinfo {author} {\bibfnamefont {V.~W.}\ \bibnamefont {Scarola}}, \ and\
  \bibinfo {author} {\bibfnamefont {D.}~\bibnamefont {Minic}},\ }\href@noop {}
  {\bibfield  {journal} {\bibinfo  {journal} {Phys. Rev. Lett.}\ }\textbf
  {\bibinfo {volume} {105}},\ \bibinfo {pages} {186801} (\bibinfo {year}
  {2010})}\BibitemShut {NoStop}%
\bibitem [{\citenamefont {Liu}\ \emph {et~al.}(2011)\citenamefont {Liu},
  \citenamefont {Peng}, \citenamefont {Tang}, \citenamefont {Sun},
  \citenamefont {Zhang},\ and\ \citenamefont {Zhong}}]{Liu2011}%
  \BibitemOpen
  \bibfield  {author} {\bibinfo {author} {\bibfnamefont {W.}~\bibnamefont
  {Liu}}, \bibinfo {author} {\bibfnamefont {X.}~\bibnamefont {Peng}}, \bibinfo
  {author} {\bibfnamefont {C.}~\bibnamefont {Tang}}, \bibinfo {author}
  {\bibfnamefont {L.}~\bibnamefont {Sun}}, \bibinfo {author} {\bibfnamefont
  {K.}~\bibnamefont {Zhang}}, \ and\ \bibinfo {author} {\bibfnamefont
  {J.}~\bibnamefont {Zhong}},\ }\href@noop {} {\bibfield  {journal} {\bibinfo
  {journal} {Phys. Rev. B}\ }\textbf {\bibinfo {volume} {84}},\ \bibinfo
  {pages} {245105} (\bibinfo {year} {2011})}\BibitemShut {NoStop}%
\bibitem [{\citenamefont {Young}\ \emph {et~al.}(2011)\citenamefont {Young},
  \citenamefont {Chowdhury}, \citenamefont {Walter}, \citenamefont {Mele},
  \citenamefont {Kane},\ and\ \citenamefont {Rappe}}]{Young2011}%
  \BibitemOpen
  \bibfield  {author} {\bibinfo {author} {\bibfnamefont {S.~M.}\ \bibnamefont
  {Young}}, \bibinfo {author} {\bibfnamefont {S.}~\bibnamefont {Chowdhury}},
  \bibinfo {author} {\bibfnamefont {E.~J.}\ \bibnamefont {Walter}}, \bibinfo
  {author} {\bibfnamefont {E.~J.}\ \bibnamefont {Mele}}, \bibinfo {author}
  {\bibfnamefont {C.~L.}\ \bibnamefont {Kane}}, \ and\ \bibinfo {author}
  {\bibfnamefont {A.~M.}\ \bibnamefont {Rappe}},\ }\href@noop {} {\bibfield
  {journal} {\bibinfo  {journal} {Phys. Rev. B}\ }\textbf {\bibinfo {volume}
  {84}},\ \bibinfo {pages} {085106} (\bibinfo {year} {2011})}\BibitemShut
  {NoStop}%
\bibitem [{\citenamefont {Zhang}\ \emph {et~al.}(2011)\citenamefont {Zhang},
  \citenamefont {Zhang}, \citenamefont {Weng}, \citenamefont {Zhang},
  \citenamefont {Yang}, \citenamefont {Liu}, \citenamefont {Feng},
  \citenamefont {Wang}, \citenamefont {Yu}, \citenamefont {Cao}, \citenamefont
  {Wang}, \citenamefont {Yang}, \citenamefont {Liu}, \citenamefont {Zhao},
  \citenamefont {Zhang}, \citenamefont {Dai}, \citenamefont {Fang},\ and\
  \citenamefont {Jin}}]{Zhang2011}%
  \BibitemOpen
  \bibfield  {author} {\bibinfo {author} {\bibfnamefont {J.~L.}\ \bibnamefont
  {Zhang}}, \bibinfo {author} {\bibfnamefont {S.~J.}\ \bibnamefont {Zhang}},
  \bibinfo {author} {\bibfnamefont {H.~M.}\ \bibnamefont {Weng}}, \bibinfo
  {author} {\bibfnamefont {W.}~\bibnamefont {Zhang}}, \bibinfo {author}
  {\bibfnamefont {L.~X.}\ \bibnamefont {Yang}}, \bibinfo {author}
  {\bibfnamefont {Q.~Q.}\ \bibnamefont {Liu}}, \bibinfo {author} {\bibfnamefont
  {S.~M.}\ \bibnamefont {Feng}}, \bibinfo {author} {\bibfnamefont {X.~C.}\
  \bibnamefont {Wang}}, \bibinfo {author} {\bibfnamefont {R.~C.}\ \bibnamefont
  {Yu}}, \bibinfo {author} {\bibfnamefont {L.~Z.}\ \bibnamefont {Cao}},
  \bibinfo {author} {\bibfnamefont {L.}~\bibnamefont {Wang}}, \bibinfo {author}
  {\bibfnamefont {W.~G.}\ \bibnamefont {Yang}}, \bibinfo {author}
  {\bibfnamefont {H.~Z.}\ \bibnamefont {Liu}}, \bibinfo {author} {\bibfnamefont
  {W.~Y.}\ \bibnamefont {Zhao}}, \bibinfo {author} {\bibfnamefont {S.~C.}\
  \bibnamefont {Zhang}}, \bibinfo {author} {\bibfnamefont {X.}~\bibnamefont
  {Dai}}, \bibinfo {author} {\bibfnamefont {Z.}~\bibnamefont {Fang}}, \ and\
  \bibinfo {author} {\bibfnamefont {C.~Q.}\ \bibnamefont {Jin}},\ }\href@noop
  {} {\bibfield  {journal} {\bibinfo  {journal} {P. Natl. Acad. Sci. USA}\
  }\textbf {\bibinfo {volume} {108}},\ \bibinfo {pages} {24} (\bibinfo {year}
  {2011})}\BibitemShut {NoStop}%
\bibitem [{\citenamefont {Lebedeva}\ \emph {et~al.}(2011)\citenamefont
  {Lebedeva}, \citenamefont {Knizhnik}, \citenamefont {Popov}, \citenamefont
  {Lozovik},\ and\ \citenamefont {Potapkin}}]{Lebedeva2011}%
  \BibitemOpen
  \bibfield  {author} {\bibinfo {author} {\bibfnamefont {I.~V.}\ \bibnamefont
  {Lebedeva}}, \bibinfo {author} {\bibfnamefont {A.~A.}\ \bibnamefont
  {Knizhnik}}, \bibinfo {author} {\bibfnamefont {A.~M.}\ \bibnamefont {Popov}},
  \bibinfo {author} {\bibfnamefont {Y.~E.}\ \bibnamefont {Lozovik}}, \ and\
  \bibinfo {author} {\bibfnamefont {B.~V.}\ \bibnamefont {Potapkin}},\
  }\href@noop {} {\bibfield  {journal} {\bibinfo  {journal} {Phys. Chem. Chem.
  Phys.}\ }\textbf {\bibinfo {volume} {13}},\ \bibinfo {pages} {5687} (\bibinfo
  {year} {2011})}\BibitemShut {NoStop}%
\bibitem [{\citenamefont {Liu}\ \emph {et~al.}(2003)\citenamefont {Liu},
  \citenamefont {Feng},\ and\ \citenamefont {Shen}}]{Liu2003}%
  \BibitemOpen
  \bibfield  {author} {\bibinfo {author} {\bibfnamefont {L.}~\bibnamefont
  {Liu}}, \bibinfo {author} {\bibfnamefont {Y.~P.}\ \bibnamefont {Feng}}, \
  and\ \bibinfo {author} {\bibfnamefont {Z.~X.}\ \bibnamefont {Shen}},\
  }\href@noop {} {\bibfield  {journal} {\bibinfo  {journal} {Phys. Rev. B}\
  }\textbf {\bibinfo {volume} {68}},\ \bibinfo {pages} {104102} (\bibinfo
  {year} {2003})}\BibitemShut {NoStop}%
\bibitem [{\citenamefont {Popov}\ \emph {et~al.}(2011)\citenamefont {Popov},
  \citenamefont {Lebedeva}, \citenamefont {Knizhnik}, \citenamefont {Lozovik},\
  and\ \citenamefont {Potapkin}}]{Popov2011}%
  \BibitemOpen
  \bibfield  {author} {\bibinfo {author} {\bibfnamefont {A.~M.}\ \bibnamefont
  {Popov}}, \bibinfo {author} {\bibfnamefont {I.~V.}\ \bibnamefont {Lebedeva}},
  \bibinfo {author} {\bibfnamefont {A.~A.}\ \bibnamefont {Knizhnik}}, \bibinfo
  {author} {\bibfnamefont {Y.~E.}\ \bibnamefont {Lozovik}}, \ and\ \bibinfo
  {author} {\bibfnamefont {B.~V.}\ \bibnamefont {Potapkin}},\ }\href@noop {}
  {\bibfield  {journal} {\bibinfo  {journal} {Phys. Rev. B}\ }\textbf {\bibinfo
  {volume} {84}},\ \bibinfo {pages} {045404} (\bibinfo {year}
  {2011})}\BibitemShut {NoStop}%
\bibitem [{\citenamefont {Kresse}\ and\ \citenamefont
  {Joubert}(1999)}]{Kresse1999}%
  \BibitemOpen
  \bibfield  {author} {\bibinfo {author} {\bibfnamefont {G.}~\bibnamefont
  {Kresse}}\ and\ \bibinfo {author} {\bibfnamefont {D.}~\bibnamefont
  {Joubert}},\ }\href@noop {} {\bibfield  {journal} {\bibinfo  {journal} {Phys.
  Rev. B}\ }\textbf {\bibinfo {volume} {59}},\ \bibinfo {pages} {1758}
  (\bibinfo {year} {1999})}\BibitemShut {NoStop}%
\bibitem [{\citenamefont {Kresse}\ and\ \citenamefont
  {Furthm\"{u}ller}(1996)}]{Kresse1996}%
  \BibitemOpen
  \bibfield  {author} {\bibinfo {author} {\bibfnamefont {G.}~\bibnamefont
  {Kresse}}\ and\ \bibinfo {author} {\bibfnamefont {J.}~\bibnamefont
  {Furthm\"{u}ller}},\ }\href@noop {} {\bibfield  {journal} {\bibinfo
  {journal} {Phys. Rev. B}\ }\textbf {\bibinfo {volume} {54}},\ \bibinfo
  {pages} {11169} (\bibinfo {year} {1996})}\BibitemShut {NoStop}%
\bibitem [{\citenamefont {Perdew}\ \emph {et~al.}(1996)\citenamefont {Perdew},
  \citenamefont {Burke},\ and\ \citenamefont {Ernzerhof}}]{Perdew1996}%
  \BibitemOpen
  \bibfield  {author} {\bibinfo {author} {\bibfnamefont {J.~P.}\ \bibnamefont
  {Perdew}}, \bibinfo {author} {\bibfnamefont {K.}~\bibnamefont {Burke}}, \
  and\ \bibinfo {author} {\bibfnamefont {M.}~\bibnamefont {Ernzerhof}},\
  }\href@noop {} {\bibfield  {journal} {\bibinfo  {journal} {Phys. Rev. Lett.}\
  }\textbf {\bibinfo {volume} {77}},\ \bibinfo {pages} {3865} (\bibinfo {year}
  {1996})}\BibitemShut {NoStop}%
\bibitem [{\citenamefont {Scheidemantel}\ \emph {et~al.}(2003)\citenamefont
  {Scheidemantel}, \citenamefont {Ambrosch-Draxl}, \citenamefont {Thonhauser},
  \citenamefont {Badding},\ and\ \citenamefont {Sofo}}]{Scheidemantel2003}%
  \BibitemOpen
  \bibfield  {author} {\bibinfo {author} {\bibfnamefont {T.~J.}\ \bibnamefont
  {Scheidemantel}}, \bibinfo {author} {\bibfnamefont {C.}~\bibnamefont
  {Ambrosch-Draxl}}, \bibinfo {author} {\bibfnamefont {T.}~\bibnamefont
  {Thonhauser}}, \bibinfo {author} {\bibfnamefont {J.~V.}\ \bibnamefont
  {Badding}}, \ and\ \bibinfo {author} {\bibfnamefont {J.~O.}\ \bibnamefont
  {Sofo}},\ }\href@noop {} {\bibfield  {journal} {\bibinfo  {journal} {Phys.
  Rev. B}\ }\textbf {\bibinfo {volume} {68}},\ \bibinfo {pages} {125210}
  (\bibinfo {year} {2003})}\BibitemShut {NoStop}%
\bibitem [{\citenamefont {Grimme}\ \emph {et~al.}(2010)\citenamefont {Grimme},
  \citenamefont {Antony}, \citenamefont {Ehrlich},\ and\ \citenamefont
  {Krieg}}]{Grimme2010}%
  \BibitemOpen
  \bibfield  {author} {\bibinfo {author} {\bibfnamefont {S.}~\bibnamefont
  {Grimme}}, \bibinfo {author} {\bibfnamefont {J.}~\bibnamefont {Antony}},
  \bibinfo {author} {\bibfnamefont {S.}~\bibnamefont {Ehrlich}}, \ and\
  \bibinfo {author} {\bibfnamefont {H.}~\bibnamefont {Krieg}},\ }\href@noop {}
  {\bibfield  {journal} {\bibinfo  {journal} {J. Chem. Phys.}\ }\textbf
  {\bibinfo {volume} {132}},\ \bibinfo {pages} {154104} (\bibinfo {year}
  {2010})}\BibitemShut {NoStop}%
\bibitem [{DFT()}]{DFTD3}%
  \BibitemOpen
  \href@noop {} {}\bibinfo {note} {Grimme's DFT-D3 program, URL: {\it
  http://toc.uni-muenster.de/DFTD3}}\BibitemShut {NoStop}%
\bibitem [{\citenamefont {Franchini}\ \emph {et~al.}(2005)\citenamefont
  {Franchini}, \citenamefont {Bayer},\ and\ \citenamefont
  {Podloucky}}]{Franchini2005}%
  \BibitemOpen
  \bibfield  {author} {\bibinfo {author} {\bibfnamefont {C.}~\bibnamefont
  {Franchini}}, \bibinfo {author} {\bibfnamefont {V.}~\bibnamefont {Bayer}}, \
  and\ \bibinfo {author} {\bibfnamefont {R.}~\bibnamefont {Podloucky}},\
  }\href@noop {} {\bibfield  {journal} {\bibinfo  {journal} {Phys. Rev. B}\
  }\textbf {\bibinfo {volume} {72}},\ \bibinfo {pages} {045132} (\bibinfo
  {year} {2005})}\BibitemShut {NoStop}%
\bibitem [{\citenamefont {Krukau}\ \emph {et~al.}(2006)\citenamefont {Krukau},
  \citenamefont {Vydrov}, \citenamefont {Izmaylov},\ and\ \citenamefont
  {Scuseria}}]{Krukau2006}%
  \BibitemOpen
  \bibfield  {author} {\bibinfo {author} {\bibfnamefont {A.~V.}\ \bibnamefont
  {Krukau}}, \bibinfo {author} {\bibfnamefont {O.~A.}\ \bibnamefont {Vydrov}},
  \bibinfo {author} {\bibfnamefont {A.~F.}\ \bibnamefont {Izmaylov}}, \ and\
  \bibinfo {author} {\bibfnamefont {G.~E.}\ \bibnamefont {Scuseria}},\
  }\href@noop {} {\bibfield  {journal} {\bibinfo  {journal} {J. Chem. Phys.}\
  }\textbf {\bibinfo {volume} {125}},\ \bibinfo {pages} {224106} (\bibinfo
  {year} {2006})}\BibitemShut {NoStop}%
\bibitem [{\citenamefont {Luo}\ \emph {et~al.}(2012)\citenamefont {Luo},
  \citenamefont {Sullivan},\ and\ \citenamefont {Quek}}]{Luo2012}%
  \BibitemOpen
  \bibfield  {author} {\bibinfo {author} {\bibfnamefont {X.}~\bibnamefont
  {Luo}}, \bibinfo {author} {\bibfnamefont {M.~B.}\ \bibnamefont {Sullivan}}, \
  and\ \bibinfo {author} {\bibfnamefont {S.~Y.}\ \bibnamefont {Quek}},\
  }\href@noop {} {\bibfield  {journal} {\bibinfo  {journal} {Phys. Rev. B}\
  }\textbf {\bibinfo {volume} {86}},\ \bibinfo {pages} {184111} (\bibinfo
  {year} {2012})}\BibitemShut {NoStop}%
\bibitem [{\citenamefont {Yazyev}\ \emph {et~al.}(2012)\citenamefont {Yazyev},
  \citenamefont {Kioupakis}, \citenamefont {Moore},\ and\ \citenamefont
  {Louie}}]{Yazyev2012}%
  \BibitemOpen
  \bibfield  {author} {\bibinfo {author} {\bibfnamefont {O.~V.}\ \bibnamefont
  {Yazyev}}, \bibinfo {author} {\bibfnamefont {E.}~\bibnamefont {Kioupakis}},
  \bibinfo {author} {\bibfnamefont {J.~E.}\ \bibnamefont {Moore}}, \ and\
  \bibinfo {author} {\bibfnamefont {S.~G.}\ \bibnamefont {Louie}},\ }\href@noop
  {} {\bibfield  {journal} {\bibinfo  {journal} {Phys. Rev. B}\ }\textbf
  {\bibinfo {volume} {85}},\ \bibinfo {pages} {161101} (\bibinfo {year}
  {2012})}\BibitemShut {NoStop}%
\bibitem [{\citenamefont {Henkelman}\ \emph {et~al.}(2000)\citenamefont
  {Henkelman}, \citenamefont {Uberuaga},\ and\ \citenamefont
  {J\'{o}nsson}}]{Henkelman2000}%
  \BibitemOpen
  \bibfield  {author} {\bibinfo {author} {\bibfnamefont {G.}~\bibnamefont
  {Henkelman}}, \bibinfo {author} {\bibfnamefont {B.~P.}\ \bibnamefont
  {Uberuaga}}, \ and\ \bibinfo {author} {\bibfnamefont {H.}~\bibnamefont
  {J\'{o}nsson}},\ }\href@noop {} {\bibfield  {journal} {\bibinfo  {journal}
  {J. Chem. Phys.}\ }\textbf {\bibinfo {volume} {113}},\ \bibinfo {pages}
  {9901} (\bibinfo {year} {2000})}\BibitemShut {NoStop}%
\bibitem [{\citenamefont {Wyckoff}(1964)}]{Wyckoff1964}%
  \BibitemOpen
  \bibinfo {editor} {\bibfnamefont {R.~W.~G.}\ \bibnamefont {Wyckoff}},\ ed.,\
  \href@noop {} {\emph {\bibinfo {title} {Crystal Structures Volume 2}}}\
  (\bibinfo  {publisher} {New York: Wiley},\ \bibinfo {year}
  {1964})\BibitemShut {NoStop}%
\bibitem [{\citenamefont {Wang}\ and\ \citenamefont {Cagin}(2007)}]{Wang2007}%
  \BibitemOpen
  \bibfield  {author} {\bibinfo {author} {\bibfnamefont {G.}~\bibnamefont
  {Wang}}\ and\ \bibinfo {author} {\bibfnamefont {T.}~\bibnamefont {Cagin}},\
  }\href@noop {} {\bibfield  {journal} {\bibinfo  {journal} {Phys. Rev. B}\
  }\textbf {\bibinfo {volume} {76}},\ \bibinfo {pages} {075201} (\bibinfo
  {year} {2007})}\BibitemShut {NoStop}%
\bibitem [{\citenamefont {Black}\ \emph {et~al.}(1957)\citenamefont {Black},
  \citenamefont {Conwell}, \citenamefont {Seigle},\ and\ \citenamefont {{C. W.
  Spencer}}}]{Black1957}%
  \BibitemOpen
  \bibfield  {author} {\bibinfo {author} {\bibfnamefont {J.}~\bibnamefont
  {Black}}, \bibinfo {author} {\bibfnamefont {E.~M.}\ \bibnamefont {Conwell}},
  \bibinfo {author} {\bibfnamefont {L.}~\bibnamefont {Seigle}}, \ and\ \bibinfo
  {author} {\bibnamefont {{C. W. Spencer}}},\ }\href@noop {} {\bibfield
  {journal} {\bibinfo  {journal} {J. Phys. Chem. Solids}\ }\textbf {\bibinfo
  {volume} {2}},\ \bibinfo {pages} {240} (\bibinfo {year} {1957})}\BibitemShut
  {NoStop}%
\bibitem [{\citenamefont {Vidal}\ \emph {et~al.}(2011)\citenamefont {Vidal},
  \citenamefont {Zhang}, \citenamefont {Yu}, \citenamefont {Luo},\ and\
  \citenamefont {Zunger}}]{Vidal2011}%
  \BibitemOpen
  \bibfield  {author} {\bibinfo {author} {\bibfnamefont {J.}~\bibnamefont
  {Vidal}}, \bibinfo {author} {\bibfnamefont {X.}~\bibnamefont {Zhang}},
  \bibinfo {author} {\bibfnamefont {L.}~\bibnamefont {Yu}}, \bibinfo {author}
  {\bibfnamefont {J.-W.}\ \bibnamefont {Luo}}, \ and\ \bibinfo {author}
  {\bibfnamefont {A.}~\bibnamefont {Zunger}},\ }\href@noop {} {\bibfield
  {journal} {\bibinfo  {journal} {Phys. Rev. B}\ }\textbf {\bibinfo {volume}
  {84}},\ \bibinfo {pages} {041109} (\bibinfo {year} {2011})}\BibitemShut
  {NoStop}%
\bibitem [{\citenamefont {Nechaev}\ \emph {et~al.}(2013)\citenamefont
  {Nechaev}, \citenamefont {Hatch}, \citenamefont {Bianchi}, \citenamefont
  {Guan}, \citenamefont {Friedrich}, \citenamefont {Aguilera}, \citenamefont
  {Mi}, \citenamefont {Iversen}, \citenamefont {BlŸgel}, \citenamefont
  {Hofmann},\ and\ \citenamefont {Chulkov}}]{Nechaev2013}%
  \BibitemOpen
  \bibfield  {author} {\bibinfo {author} {\bibfnamefont {I.~A.}\ \bibnamefont
  {Nechaev}}, \bibinfo {author} {\bibfnamefont {R.~C.}\ \bibnamefont {Hatch}},
  \bibinfo {author} {\bibfnamefont {M.}~\bibnamefont {Bianchi}}, \bibinfo
  {author} {\bibfnamefont {D.}~\bibnamefont {Guan}}, \bibinfo {author}
  {\bibfnamefont {C.}~\bibnamefont {Friedrich}}, \bibinfo {author}
  {\bibfnamefont {I.}~\bibnamefont {Aguilera}}, \bibinfo {author}
  {\bibfnamefont {J.~L.}\ \bibnamefont {Mi}}, \bibinfo {author} {\bibfnamefont
  {B.~B.}\ \bibnamefont {Iversen}}, \bibinfo {author} {\bibfnamefont
  {S.}~\bibnamefont {BlŸgel}}, \bibinfo {author} {\bibfnamefont
  {P.}~\bibnamefont {Hofmann}}, \ and\ \bibinfo {author} {\bibfnamefont
  {E.~V.}\ \bibnamefont {Chulkov}},\ }\href@noop {} {\bibfield  {journal}
  {\bibinfo  {journal} {Physical Review B}\ }\textbf {\bibinfo {volume} {87}},\
  \bibinfo {pages} {121111} (\bibinfo {year} {2013})}\BibitemShut {NoStop}%
\bibitem [{\citenamefont {Chen}\ \emph {et~al.}(2009)\citenamefont {Chen},
  \citenamefont {Analytis}, \citenamefont {Chu}, \citenamefont {Liu},
  \citenamefont {Mo}, \citenamefont {Qi}, \citenamefont {Zhang}, \citenamefont
  {Lu}, \citenamefont {Dai}, \citenamefont {Fang}, \citenamefont {Zhang},
  \citenamefont {Fisher}, \citenamefont {Hussain},\ and\ \citenamefont
  {Shen}}]{Chen2009}%
  \BibitemOpen
  \bibfield  {author} {\bibinfo {author} {\bibfnamefont {Y.~L.}\ \bibnamefont
  {Chen}}, \bibinfo {author} {\bibfnamefont {J.~G.}\ \bibnamefont {Analytis}},
  \bibinfo {author} {\bibfnamefont {J.-H.}\ \bibnamefont {Chu}}, \bibinfo
  {author} {\bibfnamefont {Z.~K.}\ \bibnamefont {Liu}}, \bibinfo {author}
  {\bibfnamefont {S.-K.}\ \bibnamefont {Mo}}, \bibinfo {author} {\bibfnamefont
  {X.~L.}\ \bibnamefont {Qi}}, \bibinfo {author} {\bibfnamefont {H.~J.}\
  \bibnamefont {Zhang}}, \bibinfo {author} {\bibfnamefont {D.~H.}\ \bibnamefont
  {Lu}}, \bibinfo {author} {\bibfnamefont {X.}~\bibnamefont {Dai}}, \bibinfo
  {author} {\bibfnamefont {Z.}~\bibnamefont {Fang}}, \bibinfo {author}
  {\bibfnamefont {S.~C.}\ \bibnamefont {Zhang}}, \bibinfo {author}
  {\bibfnamefont {I.~R.}\ \bibnamefont {Fisher}}, \bibinfo {author}
  {\bibfnamefont {Z.}~\bibnamefont {Hussain}}, \ and\ \bibinfo {author}
  {\bibfnamefont {Z.-X.}\ \bibnamefont {Shen}},\ }\href@noop {} {\bibfield
  {journal} {\bibinfo  {journal} {Science}\ }\textbf {\bibinfo {volume}
  {325}},\ \bibinfo {pages} {178} (\bibinfo {year} {2009})}\BibitemShut
  {NoStop}%
\bibitem [{\citenamefont {Sehr}\ and\ \citenamefont
  {Testardi}(1962)}]{Sehr1962}%
  \BibitemOpen
  \bibfield  {author} {\bibinfo {author} {\bibfnamefont {R.}~\bibnamefont
  {Sehr}}\ and\ \bibinfo {author} {\bibfnamefont {L.~R.}\ \bibnamefont
  {Testardi}},\ }\href@noop {} {\bibfield  {journal} {\bibinfo  {journal} {J.
  Phys. Chem. Solids}\ }\textbf {\bibinfo {volume} {23}},\ \bibinfo {pages}
  {1219} (\bibinfo {year} {1962})}\BibitemShut {NoStop}%
\bibitem [{\citenamefont {Sakamoto}\ \emph {et~al.}(2010)\citenamefont
  {Sakamoto}, \citenamefont {Hirahara}, \citenamefont {Miyazaki}, \citenamefont
  {Kimura},\ and\ \citenamefont {Hasegawa}}]{Sakamoto2010}%
  \BibitemOpen
  \bibfield  {author} {\bibinfo {author} {\bibfnamefont {Y.}~\bibnamefont
  {Sakamoto}}, \bibinfo {author} {\bibfnamefont {T.}~\bibnamefont {Hirahara}},
  \bibinfo {author} {\bibfnamefont {H.}~\bibnamefont {Miyazaki}}, \bibinfo
  {author} {\bibfnamefont {S.-i.}\ \bibnamefont {Kimura}}, \ and\ \bibinfo
  {author} {\bibfnamefont {S.}~\bibnamefont {Hasegawa}},\ }\href@noop {}
  {\bibfield  {journal} {\bibinfo  {journal} {Phys. Rev. B}\ }\textbf {\bibinfo
  {volume} {81}},\ \bibinfo {pages} {165432} (\bibinfo {year}
  {2010})}\BibitemShut {NoStop}%
\end{thebibliography}
\end{document}